%% file: main.tex
\documentclass[sigconf]{acmart}

\AtBeginDocument{%
  \providecommand\BibTeX{{%
    \normalfont B\kern-0.5em{\scshape i\kern-0.25em b}\kern-0.8em\TeX}}}

\copyrightyear{2024}
\acmYear{2024}
\setcopyright{rightsretained}
\acmConference[CHI '24]{Proceedings of the CHI Conference on Human Factors in Computing Systems}{May 11--16, 2024}{Honolulu, HI, USA}
\acmBooktitle{Proceedings of the CHI Conference on Human Factors in Computing Systems (CHI '24), May 11--16, 2024, Honolulu, HI, USA}
\acmDOI{10.1145/3613904.3642443}
\acmISBN{979-8-4007-0330-0/24/05}

\usepackage{soul}
\usepackage[prologue]{xcolor}
\newcommand{\bpstart}[1]{\noindent{\textbf{#1}}.}
\newcommand{\quotes}[1]{``#1''}

\newcommand{\DIFdel}[1]{}
\newcommand{\DIFdelFL}[1]{}
\newcommand{\DIFadd}[1]{#1}

\usepackage{fontawesome5}
\usepackage{subcaption}
\usepackage{stfloats}
\title{TutoAI: A Cross-domain Framework for AI-assisted Mixed-media Tutorial Creation on Physical Tasks}

\author{Yuexi Chen}
\authornote{Part of the work done during an internship at Adobe 
}
\email{ychen151@umd.edu}
\affiliation{%
  \institution{University of Maryland}
  \city{College Park}
  \state{Maryland}
  \country{USA}
}

\author{Vlad I. Morariu}
\email{morariu@adobe.com}
\affiliation{
\institution{Adobe Research}
  \city{College Park}
  \state{Maryland}
  \country{USA}
}
\author{Anh Truong}
\email{truong@adobe.com}
\affiliation{
\institution{Adobe Research}
 \city{San Francisco}
  \state{California}
  \country{USA}
}
\author{Zhicheng Liu}
\email{leozcliu@umd.edu}
\affiliation{%
  \institution{University of Maryland}
  \city{College Park}
  \state{Maryland}
  \country{USA}
}

\begin{document}

\begin{abstract}
Mixed-media tutorials, which integrate videos, images, text, and diagrams to teach procedural skills, offer more browsable alternatives than timeline-based videos. However, manually creating such tutorials is tedious, and existing automated solutions are often restricted to a particular domain. While AI models hold promise, it is unclear how to effectively harness their powers, given the multi-modal data involved and the vast landscape of models. We present TutoAI, a cross-domain framework for AI-assisted mixed-media tutorial creation on physical tasks. First, we distill common tutorial components by surveying existing work; then, we present an approach to identify, assemble, and evaluate AI models for component extraction; finally, we propose guidelines for designing user interfaces (UI) that support tutorial creation based on AI-generated components. We show that TutoAI has achieved higher or similar quality compared to a baseline model in preliminary user studies. 
\end{abstract}




\begin{teaserfigure}
  \includegraphics[width=\textwidth]{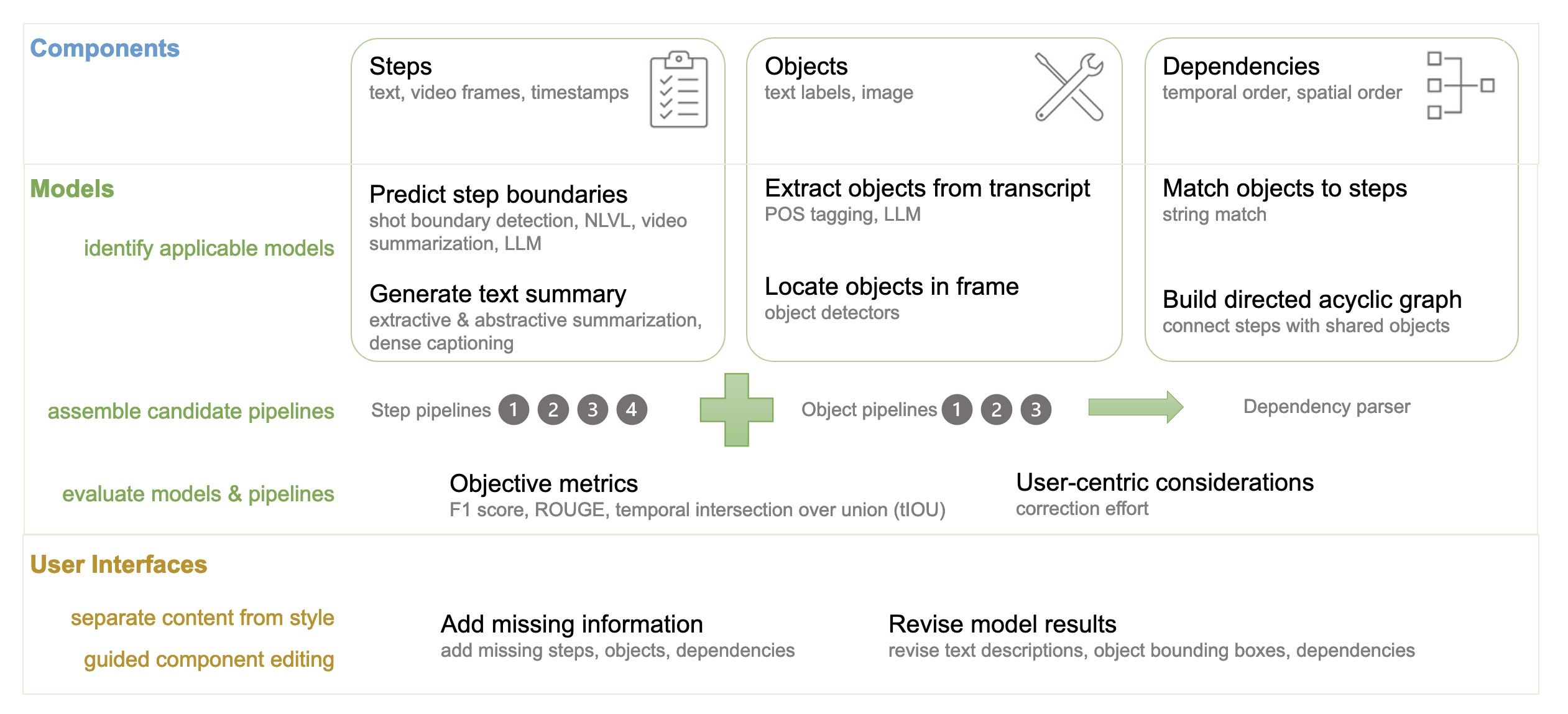}
  \caption{TutoAI is a framework for AI-assisted mixed-media tutorial creation. It has three levels: components, models, and user interfaces. After identifying components of common mixed-media tutorials, TutoAI assembles and evaluates relevant computational models to extract components. Then, it presents the results on a user interface for creators to review and edit.}
  \Description{TutoAI framework: components, models and user interfaces}
  \label{fig:framework}
\end{teaserfigure}

\maketitle
\input{1-introduction}
\input{2-related-work}

\input{3-TutoAI-framework}

\input{4-layer-1-components}

\input{5-layer-2-model}
\input{6-layer-3-UI}
\input{7-framework-evaluation}

\input{8-framework-evaluation}

\input{9-discussion}

\input{10-conclusion}

\bibliographystyle{ACM-Reference-Format}
\bibliography{sample-base}

\end{document}

%% file: 1-introduction.tex
\section{introduction}
Instructional videos are important sources for people to acquire new skills. However, the linear timeline-based video format provides limited overviews, with no explicit representation of the steps and their dependencies. Besides, navigating the timeline is tedious and imprecise. While users can fast-forward or replay videos, scrubbing the timeline might cause them to overlook vital information~\cite{zhao2022rewind, tuncer2020pause}. 

Recent work has shown that mixed-media tutorials, which unify videos, images, text, and diagrams in an interactive user interface, offer more browsable alternatives. For example, YouTube Chapters~\cite{vchapters} help navigate long-form videos: each chapter corresponds to a video segment with a short text description, a thumbnail, and a timestamp. Researchers have also proposed non-linear mixed-media tutorials for tasks such as applying makeup and cooking \cite{nawhal2019videowhiz,truong2021automatic,yang2022improving}. Such tutorials optimize user navigation by providing object details and organizing steps based on dependencies.

Although the benefits of mixed-media tutorials are confirmed, 
creating such tutorials from the original instructional videos remains challenging. \DIFadd{Current approaches for authoring mixed-media tutorials are usually domain-specific, with both the tutorial components and extraction techniques tailored for each domain~\cite{fraser2020temporal, nawhal2019videowhiz, chi2012mixt, truong2021automatic}.}
\DIFadd{While many have acknowledged the importance of generalization and argued how their approaches could apply to tutorials in other domains \cite{truong2021automatic,kim2014crowdsourcing, weir2015learnersourcing, wang2021soloist},}
\DIFadd{a cross-domain framework with shared vocabulary and reusable methodologies for mixed-media tutorial creation is still lacking.} \DIFadd{We believe such a framework will 
benefit the future development of mixed-media tutorial creation, as demonstrated in other research areas~\mbox{
\cite{blackwell2003notational, calvary2003unifying}}\hskip0pt
.
} 

Recent advances in AI, especially large language models (LLM)~\cite{brown2020language},  have shown promise in content understanding and generation, \DIFadd{and can potentially play a vital role in establishing a cross-domain framework}. However, integrating AI with mixed-media tutorial creation is not straightforward. First, we have neither a vocabulary to describe the common components of mixed-media tutorials nor a systematic account of the roles of humans and AI in extracting such components. Second, a single component may have multi-modal appearances (e.g., cooking ingredients appearing in both the audio narration and video frames), and multiple machine learning (ML) models are applicable. Currently, there are no guidelines on how to assemble and evaluate ML models to obtain mixed-media tutorial components from original videos. Though the landscape of ML models  \DIFdel{change }\DIFadd{changes} over time, we believe there are general guidelines that could transcend specific models.  

To address these challenges, we present TutoAI, the first cross-domain framework to integrate AI in creating mixed-media tutorials (Figure~\ref{fig:framework}). We focus on instructional videos on physical tasks (e.g., cooking, hardware assembly) instead of concepts (e.g., lectures) or digital artifacts (e.g., software usage, programming). The TutoAI framework has three levels: components, models, and user interfaces (UI). At the \textit{component} level, we conduct a comprehensive survey to identify common components of mixed-media tutorials and analyze their representations. At the \textit{model} level, we review ML methods to extract each component and present an approach to assemble and evaluate applicable ML models. At the \textit{UI} level, we propose guidelines for building UIs that allow creators to review and edit AI-generated components and also implement an example interactive prototype. 

We evaluate TutoAI in two ways. At the model level, we validate the performance of the assembled ML pipeline on a large set of cooking videos and a small set of diverse instructional videos. At the UI level, we evaluate the user-perceived component quality by conducting two studies with 24 general instructional video viewers and 2 YouTube creators. Our results show that TutoAI-generated components have higher or similar quality compared to a baseline model (YouTube Chapters~\cite{vchapters}), and the TutoAI framework has the potential to be integrated into creators' workflow. In summary, we make the following contributions: 

\begin{itemize}
    \item A comprehensive survey for mixed-media tutorials and a taxonomy of mixed-media tutorial components.
    \item TutoAI, a cross-domain framework for AI-assisted mixed-media tutorial creation on physical tasks, including components, models, and UIs.
    \item Empirical evaluation of TutoAI framework in terms of model quality, user-perceived quality, and workflow integration.
\end{itemize}



%% file: 2-related-work.tex
\section{Related work}
\subsection{Mixed-media tutorials}
Mixed-media tutorials, \DIFadd{though diverse in format}\DIFadd{, share commonalities in tutorial components and extraction methods}. 

\DIFadd{{\bpstart{Tutorial components}}}
A common component is \DIFadd{a step, usually a video segment}, comprising a text description, a thumbnail, and a timestamp~\cite{vchapters, kim2014crowdsourcing, fraser2020temporal, pavel2014video, weir2015learnersourcing}. \DIFadd{A step could range from a cooking procedure~\mbox{
\cite{chang2018recipescape} }
to a software operation~\mbox{
\cite{fraser2020temporal}}
}. Another common component is objects, e.g., ingredients and equipment for cooking tutorials~\cite{nawhal2019videowhiz, yang2022improving, allRecipes}\DIFadd{. Besides steps and objects, some tutorials also organize steps based on dependencies, e.g., Truong et al. ~\mbox{
\cite{truong2021automatic} }
grouped makeup video segments by facial parts in a two-level hierarchical format; Nawhal et al.~\mbox{
\cite{nawhal2019videowhiz} }
and Yang et al.~\mbox{
\cite{yang2022improving} }
arranged cooking steps non-linearly by temporal and spatial dependencies. TutoAI, our proposed framework, has a \textit{Components} level built upon components distilled from existing mixed-media tutorials. 
}

\DIFadd{~\mbox{\bpstart{Extraction methods}}}
\DIFadd{Tutorial component extraction from original videos could be manual},\DIFdel{UI widgets for software tutorials ~\mbox{
\cite{fraser2020temporal}}
, or abstract concepts for lecture videos~\mbox{
\cite{liu2018conceptscape}}
. Extracting video segments and object information could be purely manual~\mbox{
\cite{allRecipes, yang2022improving, wikihow, weir2015learnersourcing}}
,} \DIFadd{automatic, or mixed-initiative (detailed comparison in Appendix Table 2-4).}\DIFdel{, with human annotators providing } \DIFadd{Websites like WikiHow~\mbox{
\cite{wikihow} }
and Allrecipes~\mbox{
\cite{allRecipes} }
depend on experts to draft tutorials; Crowdy~\mbox{
\cite{weir2015learnersourcing} }
requires learners to identify subgoals and steps. In certain domains, automatic extraction methods are feasible. MixT~\cite{chi2012mixt}
segments PhotoShop videos using software logs. Fraser et al. ~\cite{fraser2020temporal} implement a dynamic programming method to segment creative stream videos based on the transcript and software logs; Truong et al.~\cite{truong2021automatic} apply video shot detection and transcript segmentation methods for makeup videos. However, the above methods require domain-specific data and may not apply to other domains. Mixed-initiative methods involve both human effort and computational techniques. Humans could provide} input\DIFdel{~\mbox{
\cite{kim2014crowdsourcing, wang2014evertutor} }
or refining the computational results~\mbox{
\cite{nawhal2019videowhiz, chang2018recipescape, pavel2014video}}
, or fully automatic for  domains like software~\mbox{
\cite{fraser2020temporal, chi2012mixt, truong2021automatic}}
. }
\DIFdel{The TutoAI framework }, e.g., ToolScape~\cite{kim2014crowdsourcing} gathers steps from crowdworkers and converges them through clustering algorithms. EverTutor~\cite{wang2014evertutor} converts smartphone demonstrations by humans into interactive tutorials. Humans could also refine computational results, e.g., VideoWhiz~\cite{nawhal2019videowhiz} 
and RecipeDeck~\cite{chang2018recipescape} both employ Part-of-Speech (POS) tagging to detect cooking actions and objects and then rely on annotators to refine the results. Video Digests~\cite{pavel2014video}
applies Bayesian topic segmentation to generate chapters in lecture videos, allowing users to improve upon them. The second level of TutoAI focuses on models, including an approach to identifying, evaluating, and assembling AI models to extract tutorial components. TutoAI also adopts a mixed-initiative approach, where humans refine computational results.

Cross-domain applicability is a goal in previous work on mixed-media tutorials. For example, Truong et al. suggest their segmentation algorithm for makeup videos could be adapted for cooking, DIY, and bartending~\cite{truong2021automatic}; Soloist~\cite{wang2021soloist} transforms instructional guitar videos into mixed-media tutorials, and the processing pipeline can be generalized to other instruments; Kim et al. show that the same annotation pattern combined with a clustering algorithm can process cross-domain instructional videos~
\cite{kim2014crowdsourcing}; Crowdy~\cite{weir2015learnersourcing} is a subgoal-based crowdsourcing annotation workflow. 

TutoAI extends this line of work, aiming to create a general cross-domain framework for mixed-media tutorials. Unlike crowdsourcing annotation workflows, TutoAI relies on AI.\enlargethispage{12pt}  

\subsection{AI-assisted creation}
AI has augmented human creativity, from generating visuals~\cite{ramesh2021zero, midjourney} to crafting slogans and aiding scientific writing\DIFdel{with LLMs} ~\cite{brown2020language, gero2022sparks}. However, AI outputs \DIFadd{may be} imperfect or misaligned with user intentions, necessitating human refinement. \DIFdel{Tools like } \DIFadd{Researchers have built AI-assisted creation tools in multiple domains, e.g.,} Cococo~\cite{louie2020novice} allows users to adjust the mood of AI-generated music notes. \DIFdel{Dang et al.'s text editor\mbox{
\cite{dang2022beyond} }
supports user refinement of automatically generated paragraph summaries.} Morai Maker~\cite{guzdial2019friend} is a game-level editor in which human and AI designers take turns to build a Super Mario Bros game. LaMPost~\cite{goodman2022lampost} facilitates email writing for people with Dyslexia. \DIFdel{Unlike previous work focusing on a single modality, TutoAI supports multi-modal mixed-media tutorial creation empowered by various ML models. } \DIFadd{Dang et al.'s text editor~\mbox{
\cite{dang2022beyond}} supports writers to refine automatically generated paragraph summaries. Some tools focus on refinement instead of creation: e.g.,} refinement of topics returned by topic models~\cite{smith2018closing}; repair of auto-extracted PDF tables~\cite{hoffswell2019interactive}; refinement of medical images retrieved by ML models~\cite{cai2019human}. \DIFadd{TutoAI also adopts an AI-assisted approach, supporting the creation of mixed-media tutorials with extensive refinement. Unlike previous work focusing on a single modality, TutoAI supports multi-modal mixed-media tutorial creation empowered by various ML models.}

\DIFadd{Providing guardrails for AI output is crucial. Previous research has} proposed several principles for designing such mixed-initiative user interfaces ~\cite{horvitz1999principles, amershi2019guidelines}, such as \quotes{provide mechanisms for efficient agent-user collaboration to refine results} and \quotes{support efficient correction}. TutoAI adheres to these principles, and additionally shares design considerations for choosing ML methods across modalities.

\subsection{\DIFadd{Large language models (LLM) prompting}}
\DIFadd{Large language models (LLM)~\mbox{
\cite{brown2020language, thoppilan2022lamda, bommasani2021opportunities}}
, trained on internet-scale data, have demonstrated extraordinary potential in information processing tasks such as text summarization. Users interact with LLMs by providing natural language descriptions of the task, also called }\textit{\DIFadd{prompting}}\DIFadd{~\mbox{
\cite{prompt}}
. The most commonly used prompting technique is zero-shot prompting~\mbox{
\cite{betz2021thinking}}
, which describes the task directly. There are also other prompting techniques, including few-shot prompting~\mbox{
\cite{lu2021fantastically} }
and prompt chaining~\mbox{
\cite{wu2022ai}}
. Researchers have applied zero-shot prompting to summarize various types of data, including news~\mbox{
\cite{goyal2022news, zhang2023benchmarking}}
, Reddit posts~\mbox{
\cite{yang2023exploring}}
, meeting records~\mbox{
\cite{laskar2023building} }
and stories~\mbox{
\cite{yang2023exploring}}
. Researchers have also applied LLMs to summarize video transcripts. Croitoru et al. ~\mbox{
\cite{croitoru2023moment} }
applied GPT-3 to summarize software tutorial video transcripts and then used the summary to detect key moments. LUSE~\mbox{
\cite{shangluse} }
also uses zero-shot prompting to summarize tutorial video transcripts and generalize steps for a task across different videos. To evaluate the summarization quality of LLMs, researchers have used traditional metrics like ROUGE scores~\cite{lin2004rouge}, which measures the number of overlapped n-grams in the reference and summarized text, as well as employed humans to examine different aspects of the output, including coverage~\cite{shangluse}, descriptivity~\cite{shangluse}, coherence~\cite{zhang2023benchmarking}, faithfulness~\cite{zhang2023benchmarking}, relevance~\cite{zhang2023benchmarking} and personal preferences~\cite{goyal2022news}.}

\DIFadd{TutoAI also relies on zero-shot prompting to summarize video transcripts. In addition to requesting a summary, TutoAI also asks an LLM to extract objects and timestamp information. Like Croitoru et al. ~\mbox{
\cite{croitoru2023moment} }
and LUSE~\mbox{
\cite{shangluse}}
, TutoAI uses the generated summary as input for other models. The difference is that their contributions are models that focus on a single task (e.g., detect video moments) and exclude humans from the loop, but TutoAI contributes an AI-assisted framework. As LLMs suffer from }\textit{\DIFadd{hallucination}} \DIFadd{(plausible yet incorrect output)~\mbox{
\cite{zhang2023siren}}
, involving human refinement is crucial for end users. Similar to previous research, we manually evaluated the output besides ROUGE scores.
}

%% file: 3-TutoAI-framework.tex
\section{TutoAI Overview: an AI-assisted framework}
The TutoAI framework aims to provide a cross-domain approach to AI-assisted creation of mixed-media tutorials on physical tasks. We expect the input to include an instructional video and its transcript. Our design goals, informed by the review of current mixed-media tutorials and ML methods, are:

\enlargethispage{12pt}
\begin{itemize}
    \item[D1] \textbf{Support cross-domain tutorial creation:} 
    Mixed-media tutorials are useful in diverse domains, and TutoAI should offer a generalized approach. 
    \item[D2] \textbf{Handle multi-modal data types: } The input instructional videos and the output mixed-media tutorials both contain multi-modal data. TutoAI should support multi-modality.
    \item[D3] \textbf{Empower creators without information overload:} Given the multi-modalities in mixed-media tutorials and the vast landscape of ML models, TutoAI should present information to creators without overwhelming them.
\end{itemize}

\subsection{Level 1: Components}
As shown in Figure~\ref{fig:framework}, TutoAI \DIFdel{builds} \DIFadd{is built} on three \DIFadd{types of} cross-domain components in mixed-media tutorials (\textbf{D1}): steps, objects, and dependencies (detailed in section 4). These components are multi-modal (\textbf{D2}), specifically:

\begin{itemize}
    \item \textit{Steps:} represented as text, images, video clips, and temporal metadata (timestamps)
    \item  \textit{Objects:} represented as text, images, and temporal metadata (appearance time in videos)
    \item  \textit{Dependencies:} encoded as hierarchical structures, diagrams, and links

\end{itemize}

The output mixed-media tutorials may include all or a subset of these components. For instance, YouTube Chapters~\cite{vchapters} only utilize \textit{steps}. 
 \DIFdel{While objects might be mentioned in steps, YouTube Chapters do not extract nor represent objects explicitly. } 
For completeness (\textbf{D2}), we discuss all \DIFadd{three component types} in level 2 and level 3. 

\subsection{Level 2: Models}

After identifying the components and their representations, \DIFadd{we} focus on methodologies to select and evaluate applicable ML models to obtain such components from instructional videos. 
Even though cutting-edge ML models change over time, the general approaches we suggest here transcend \DIFdel{the particular models we've considered}\DIFadd{particular models} (Section 5).


\subsubsection{Identifying relevant models}
The first task is identifying models capable of extracting information required for a component. We consider models that take visual or transcript data from the video as inputs (\textbf{D2}), and with outputs that match the desired component representations. For instance, if a step component requires text descriptions, then  \DIFdel{both text summarization and dense captioning models are applicable: text summarization models process video transcripts and dense captioning models process video}  \DIFadd{models that ingest video transcripts or} frames, and \DIFdel{both produce text descriptions.
}\DIFadd{output text descriptions are applicable.} 


\subsubsection{Assembling models}
After identifying relevant models, we assemble models into candidate pipelines based on input and output modalities. For example, if a step component requires text descriptions and timestamps, instead of finding a single model that generates both, we can assemble two different pipelines serving the same goal. \DIFdel{text summarization with Natural Language Video Localization (NLVL) models, as text summarization offers text descriptions that NLVL models use to locate timestamps} \DIFadd{In the first pipeline, one model generates text descriptions, and the other locates the descriptions} in the video.
\DIFadd{Alternatively, we can assemble another pipeline where one model segments videos first and the other model generates text descriptions for each segment.}
 \DIFdel{we can assemble shot boundary detection and text summarization models. First, we obtain timestamps by shot boundary detection and then generate text summarization within boundaries.} 


\subsubsection{Evaluating models}
After considering alternative ways to assemble models, we first find common benchmark metrics \DIFadd{for model evaluation}. Besides objective metrics, we also assess correction efforts for creators.  \DIFdel{Pipelines with poor-performing models are discarded to prevent error accumulation.} For example, false \DIFadd{positives (FPs)} are deemed easier to  \DIFdel{correct than false negative (FN)errors, as FP errors only require}\DIFadd{fix than false negatives (FNs), as fixing FPs requires} deletion, but\DIFdel{FN errors require content creation from scratch.
} \DIFadd{fixing FNs requires creation.} 




\subsection{Level 3: User Interface (UI) design}

AI-generated results are typically imperfect, requiring further refinement from humans. As shown in Figure~\ref{fig:framework}, the UI should support creators to review \DIFdel{AI-generated results, add missing information, and revise and delete results.}\DIFadd{and revise AI-generated results.} To manage cognitive load (\textbf{D3})\DIFdel{and mitigate error propagation in pipelines}, the UI should display AI-generated results sequentially, allowing creators to focus on\DIFdel{refining } one aspect at a time, and the refined \DIFdel{output}\DIFadd{results} could be input for subsequent stages, \DIFadd{mitigating error propagation}. Section \ref{sec:ui} discusses UI design guidelines and presents an example implementation.


%% file: 4-layer-1-components.tex
\section{Level 1: components in mixed-media tutorials}
To explore the design space of cross-domain mixed-media tutorials, we analyzed 13 mixed-media tutorials from three websites~\mbox{\cite{wikihow, youTube, allRecipes}} and 10 research papers~{\cite{chi2012mixt, pavel2014video, wang2014evertutor, kim2014crowdsourcing, liu2018conceptscape, chang2018recipescape, nawhal2019videowhiz, fraser2020temporal, truong2021automatic, yang2022improving}}, covering at least five domains including cooking, makeup, vehicle repair, software usage, and educational lectures. Though we focus on tutorials of physical tasks, we also borrow inspiration from other domains, e.g., lectures.

For each mixed-media tutorial, we annotated the informational units, such as ingredients in recipe tutorials, and visual representations. These units were then categorized into three types of components: step, object, and dependency. \DIFadd{We also annotated extraction methods based on human roles (Appendix Table 2-4)}. 




\subsection{Step} 

Every tutorial comprises a sequence of steps, e.g., ``duplicating a layer'' in a PhotoShop tutorial ~\mbox{\cite{chi2012mixt}}. These steps may be conveyed through text, images, and video clips. Among the 13 tutorials we studied, 12 used text descriptions, 10 featured images, and 7 included video clips. Auxiliary elements can enrich the primary media. Timestamps help locate the step in the original video, overlays emphasize parts of an image, and glyphs connect images or text. We found 5 out of 7 tutorials with video clips also provide timestamps; two tutorials have overlays on images, and one uses glyphs.


Figure~\ref{fig:steps} provides step examples in mixed-media tutorials. Specifically, Figure~\ref{fig:step-a} shows a step in an interactive smartphone tutorial, marked by an overlay indicating the screen area to be clicked~\mbox{\cite{wang2014evertutor}}; Figure~\ref{fig:step-b} depicts an auto-generated YouTube Chapter for a DIY craft video featuring text, images, and video clips (with timestamps); Figure~\ref{fig:step-c} illustrates a step in a cooking tutorial, where red and blue dots signify ingredients and actions, respectively~\mbox{\cite{chang2018recipescape}}. Comprehensive details are in the Appendix.

\newsavebox{\arrangebox}
\newlength{\arrangeht}
\begin{figure}[hbtp]
\centering

\sbox{\arrangebox}{%
  \begin{subfigure}[b]{0.4\columnwidth}
  \centering
  \includegraphics[width=\textwidth]{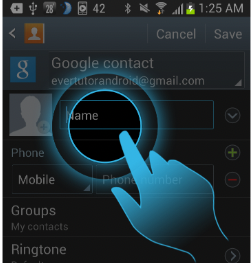}
  \caption{A step with an image and overlays in a smartphone tutorial ~\mbox{[69]}}
  \label{fig:step-a}
  \end{subfigure}%
}
\setlength{\arrangeht}{\ht\arrangebox}

\usebox{\arrangebox}\hfill
\begin{minipage}[b][140pt][s]{0.5\columnwidth}
  \begin{subfigure}[t]{\textwidth}
  \centering
  \includegraphics[width=\textwidth]{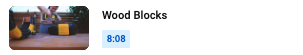}
  \caption{A step with text, images, video clips, and temporal metadata in a DIY craft tutorial~\mbox{[44]}}
  \label{fig:step-b}
  \end{subfigure}\vfill
  \begin{subfigure}[b]{\textwidth}
  \centering
  \includegraphics[width=\textwidth]{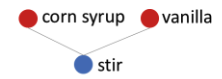}
  \caption{A step with text and glyph in a cooking tutorial~\mbox{[12]}}
  \label{fig:step-c}
  \end{subfigure}
\end{minipage}
\caption{Examples of steps in mixed-media tutorials (images used with permission)}
\label{fig:steps}\vspace*{-9pt}
\end{figure}

\subsection{Object} 

Many mixed-media tutorials explicitly specify objects required for the task, such as ingredients and equipment in cooking tutorials~\mbox{\cite{yang2022improving}}, and UI widgets in software tutorials~\mbox{\cite{fraser2020temporal}}. 
These objects can be represented through text, images, and timestamps marking their appearance in videos. In our dataset of 13 mixed-media tutorials, 7 explicitly included object components. While the remaining 6 tutorials contained objects implicitly in the instructions, they did not extract and represent these objects as individual components.
All 7 tutorials with object components featured text descriptions, 3 incorporated object images, and 2 had appearance time in videos. Additionally, 3 offered interaction features, including checkboxes or clickable buttons that link objects with other components. 

\begin{figure}
  \centering
  \begin{subfigure}{0.45\textwidth}
    \centering
    \includegraphics[width=0.45\textwidth]{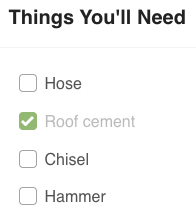}
    \caption{Object components represented using text and interactive check-boxes in a roof repairing tutorial~\mbox{[6]}}
    \label{fig:obj-a}
  \end{subfigure}%
  \hspace{5mm}
  \begin{subfigure}{0.45\textwidth}
    \centering
    \includegraphics[width=0.7\textwidth]{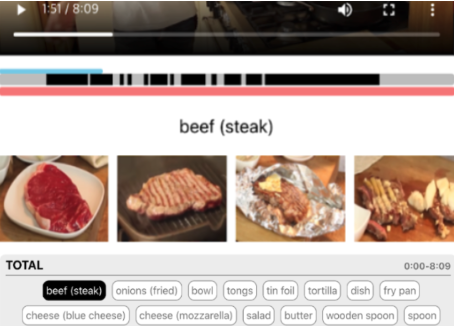}
    \caption{Object components represented using text, images, and appearance time in a cooking tutorial~\mbox{[75]}}
    \label{fig:obj-b}
  \end{subfigure}
  \caption{Examples of objects in mixed-media tutorials (images used with permission).}
  \label{fig:objects}
\end{figure}

Figure~\ref{fig:objects} illustrates examples of objects in mixed-media tutorials. Figure~\ref{fig:obj-a} displays an object component from a roof repair tutorial on WikiHow~\mbox{\cite{repairRoof}}, with interactive checkboxes to help users gather things needed; Figure~\ref{fig:obj-b} shows object buttons; clicking on an object button (e.g., ``beef (steak)'') brings up video frames containing that object and the appearance time on the timeline~\mbox{\cite{yang2022improving}}. All the examples are in the Appendix.

\begin{figure}[hbtp]
\centering
\begin{subfigure}{0.4\textwidth}
  \centering
  \includegraphics[width=0.7\textwidth]{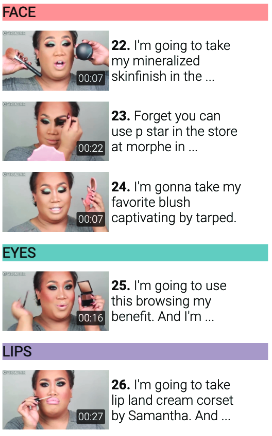}
  \caption{Spatial dependencies in a makeup tutorial~\mbox{[66]}}
  \label{fig:dep-a}
  \end{subfigure}%
\\
  \begin{subfigure}{0.4\textwidth}
  \centering
  \includegraphics[width=0.8\textwidth]{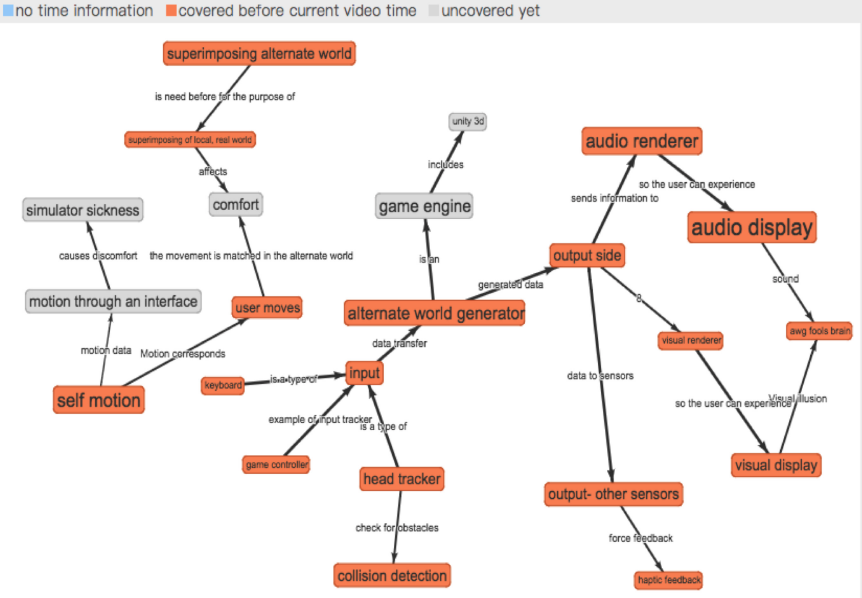}
  \caption{Concept prerequisites in a lecture~\mbox{[39]}}
  \label{fig:dep-b}
  \end{subfigure}
\\
  \begin{subfigure}{0.4\textwidth}
  \centering
  \includegraphics[width=0.8\textwidth]{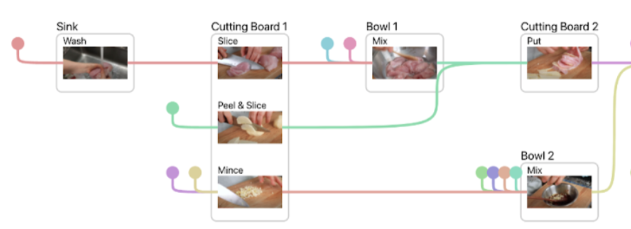}
  \caption{Action dependencies in a cooking tutorial~\mbox{[75]}}
  \label{fig:dep-c}
  \end{subfigure}

\caption{Dependency examples in mixed-media tutorials (images used with permission). 
}
\label{fig:dependencies}
\end{figure}

\DIFdel{Similarly, we group object component extraction by human involvement in Table.}

\DIFdel{Tutorial websites like WikiHow and AllRecipes rely on experts to curate objects. Identifying concepts in lectures requires students to sift through videos manually.} 
 
\DIFdel{When additional domain-specific information is available, such as software logs, objects can be automatically identified, however, these automated methods may not apply to other domains.} 

\DIFdel{Humans could refine object extraction results. For instance, POS tagging can detect nouns in text, which are candidate objects for human refinement. However, if the results contain too many irrelevant objects, the annotators may dismiss the computational results altogether and start from scratch instead.}

\subsection{Dependency}
Dependencies between steps are everywhere; they could be food processing order in recipe tutorials~\mbox{\cite{nawhal2019videowhiz, chang2018recipescape, yang2022improving}}, concept prerequisites in lectures~\mbox{\cite{liu2018conceptscape}} and facial parts in makeup tutorials~\mbox{\cite{truong2021automatic}}. Dependencies may imply a different order than the one presented in the original instructional video. For example, in a cake recipe video, though the preparations of dry and wet ingredients are shown sequentially, they could be done in parallel~\mbox{\cite{weir2015learnersourcing}}. In the TutoAI framework, we focus on physical tasks, where the dependencies between steps are the execution order. Of our collected 13 examples, 5 include dependencies explicitly. Among those 5, 4 utilize spatial layout to encode the dependency, 3 have links in the diagram. 

Figure~\ref{fig:dependencies} shows dependency examples. Figure~\ref{fig:dep-a} shows groupings in a makeup tutorial where steps within each group are sequential but independent of other groups. Figure~\ref{fig:dep-b} maps out the dependencies of concepts in a lecture: orange nodes are already covered, and gray nodes are not.  Figure~\ref{fig:dep-c} outlines cooking steps in different rows and columns: steps on the same row must be done sequentially, but steps on different rows could be done simultaneously; steps are also grouped by spatial dependencies (e.g., cutting board) in rectangles. All the examples are in the Appendix.

%% file: 5-layer-2-model.tex
\section{Level 2: Assemble and evaluate models}
We first review applicable models and candidate pipelines to extract mixed-media tutorial components. 
We then evaluate them on an annotated dataset of 347 cooking videos and finalize a pipeline. Note that we only apply ML models to step and object extraction; for dependencies, we build a directed acyclic graph (DAG) based on the temporal order and shared objects between steps. 

\subsection{Applicable models and candidate pipelines}
\subsubsection{Step extraction}
For the sake of completeness, we assume that a step component needs the following: a text description, the start and end timestamps in the video, and a representative video frame (thumbnail). As mentioned in section 3.2, we first identify relevant models: 

\begin{itemize}
    \item \bpstart{Models for text descriptions}
We identified two types of models for generating text descriptions: text summarization and video dense captioning. Text summarization takes a chunk of text as input and shortens it while preserving the key information\DIFadd{~\mbox{
\cite{erkan2004lexrank, mihalcea2004textrank, lewis2019bart, raffel2020exploring}}
.} Video dense captioning takes video frames and step timestamps as input and generates text descriptions for objects and their interactions within the step's boundary~\cite{johnson2016densecap, zhou2018end, wang2021end}.

\item \bpstart{Models for step timestamps} We identified four model types for obtaining step timestamps: natural language video localization (NLVL), shot boundary detection, video summarization, and LLM prompting. NLVL localizes the start and end time of\DIFdel{that step in the video} \DIFadd{a step given a video and a step text description}~\cite{gao2017tall, zhang2019man, rodriguez2021dori}. Shot boundary detection takes video frames as input, and returns candidate shot transition frames. Assuming that each shot represents a step, we can convert adjacent transition frame indices into the start and end timestamps~\cite{souvcek2020transnet, zhou2018towards}. Video summarization condenses a long video by selecting and stitching together keyframes to form a shorter video~\DIFadd{\mbox{
\cite{videoSummarization, zhu2020dsnet, he2023align, song2015tvsum}}
.} Similar to shot boundary detection, we can convert adjacent keyframe indices into step timestamps. We can also prompt LLMs to generate step timestamps if the input transcript contains word or sentence-level timestamps.

\item \bpstart{Models for step thumbnails} We identified two types of models for selecting thumbnails: video summarization and shot boundary detection. As mentioned before, video summarization outputs representative keyframes. In addition to representative keyframes, shot boundary detection can filter dissimilar frames to get more thumbnail candidates.

\end{itemize}

\begin{figure*}
    \centering
    \includegraphics[width=\textwidth]{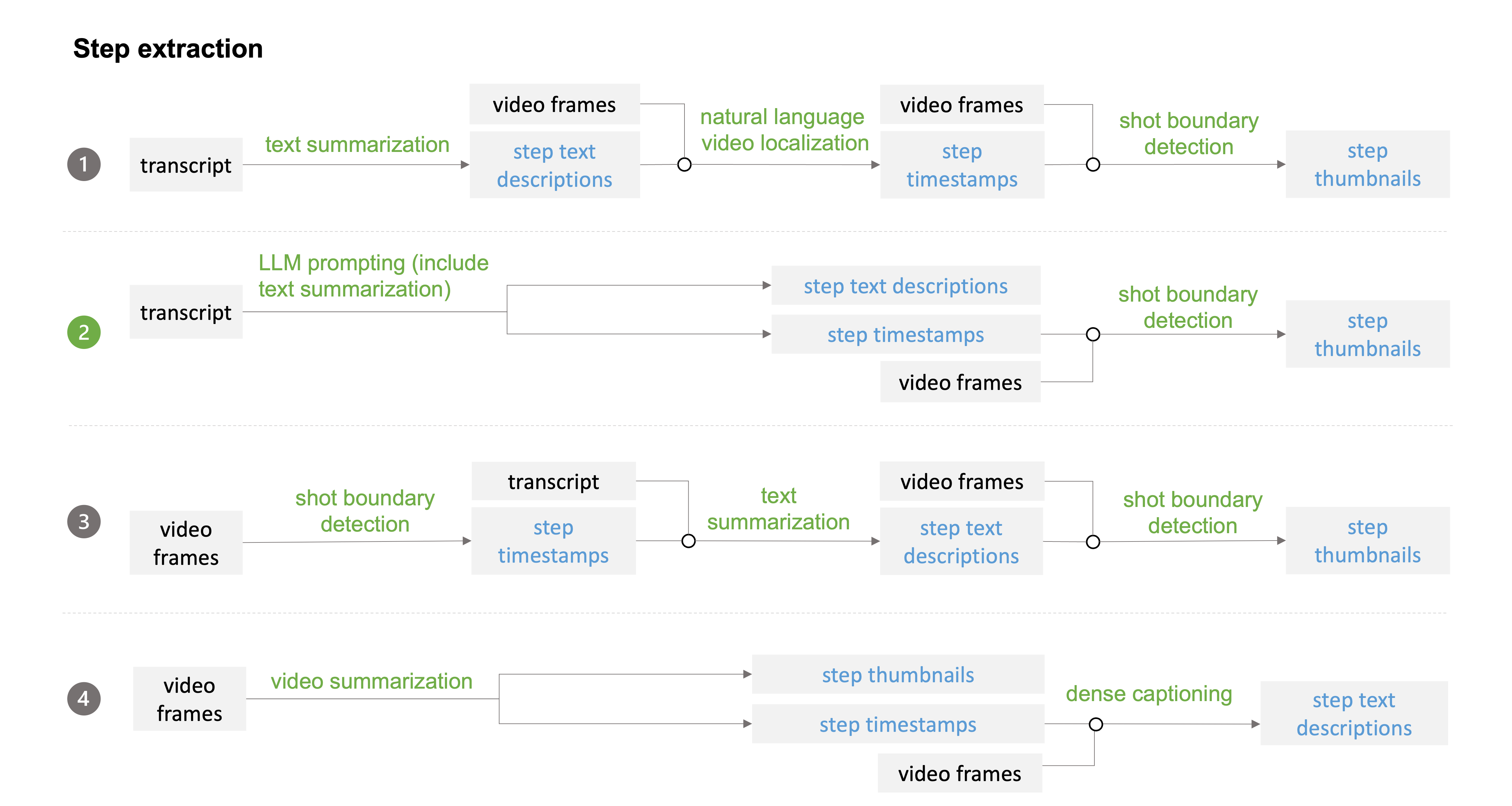}
    \caption{Four candidate pipelines for step extraction. Models are in green, and generated subcomponents are in blue. After evaluation, the chosen one is No.2.}
    \label{fig:step-extraction}
\end{figure*}

To assemble pipelines that extract all the step information, we start with models that take video frames and transcripts as input and chain additional models based on the output. Figure~\ref{fig:step-extraction} shows 4 candidate pipelines.

\begin{itemize}
    \item \bpstart{Pipeline 1: text summarization + NLVL + shot boundary detection}
As shown in Figure~\ref{fig:step-extraction}, pipeline (1) uses text summarization to extract step descriptions from the transcripts. Using step descriptions and the input video frames, it then leverages NLVL to obtain step timestamps. Lastly, it applies shot boundary detectors  
to derive thumbnails.
  \item \bpstart{Pipeline 2: LLM + shot boundary detection} Pipeline (2) uses LLM prompting to fetch both step descriptions and timestamps, followed by shot boundary detection to produce step thumbnails.
  \item \bpstart{Pipeline 3: shot boundary detection + text summarization + shot boundary detection}
 \DIFadd{Pipeline} (3) begins with shot boundary detection to obtain step timestamps, followed by text summarization for text descriptions of each step, and concludes with another round of shot boundary detection for step thumbnails.
 \item \bpstart{Pipeline 4: video summarization + video dense captioning} \DIFadd{Pipeline} (4) employs video summarization to identify step thumbnails, and then obtains timestamps by converting adjacent keyframe indices into start and end timestamps. Given timestamps and video frames, dense captioning models generate step descriptions.
\end{itemize}

\subsubsection{Object extraction}
For the sake of completeness, we assume that an object component needs the following information: object names and an image containing the \DIFdel{bounding box of the object. Similar to step extraction, we first identify}\DIFadd{object's bounding box. We have identified} relevant models:

\begin{itemize}
    \item \bpstart{Models for object names}
We identified three types of models to extract object names: Part-of-Speech (POS) taggers, LLM prompting, and traditional object detectors. POS taggers take text as input, categorizing words' roles in a sentence with grammatical properties such as nouns and verbs~\cite{posTag}. \DIFadd{Obtaining object names} from POS tagging results requires parsing nouns. LLMs can also be prompted to extract object names from text input. Traditional object detectors are trained on predefined object categories and, given input images, output detection names and bounding boxes~\cite{lin2014microsoft, fang2021you}.

\item \bpstart{Models for object bounding boxes}
We identified two types of models to obtain object bounding boxes: traditional and open-vocabulary object detectors. As mentioned before, traditional object detectors take images as input and return bounding boxes as output. However, it can only recognize objects in the training dataset. Open-vocabulary object detectors take in both object names and images, and output bounding boxes for the object names~\cite{kamath2021mdetr, minderer2022simple, li2022grounded}.
\end{itemize}

\begin{figure*}
    \centering
    \includegraphics[width=0.85\textwidth]{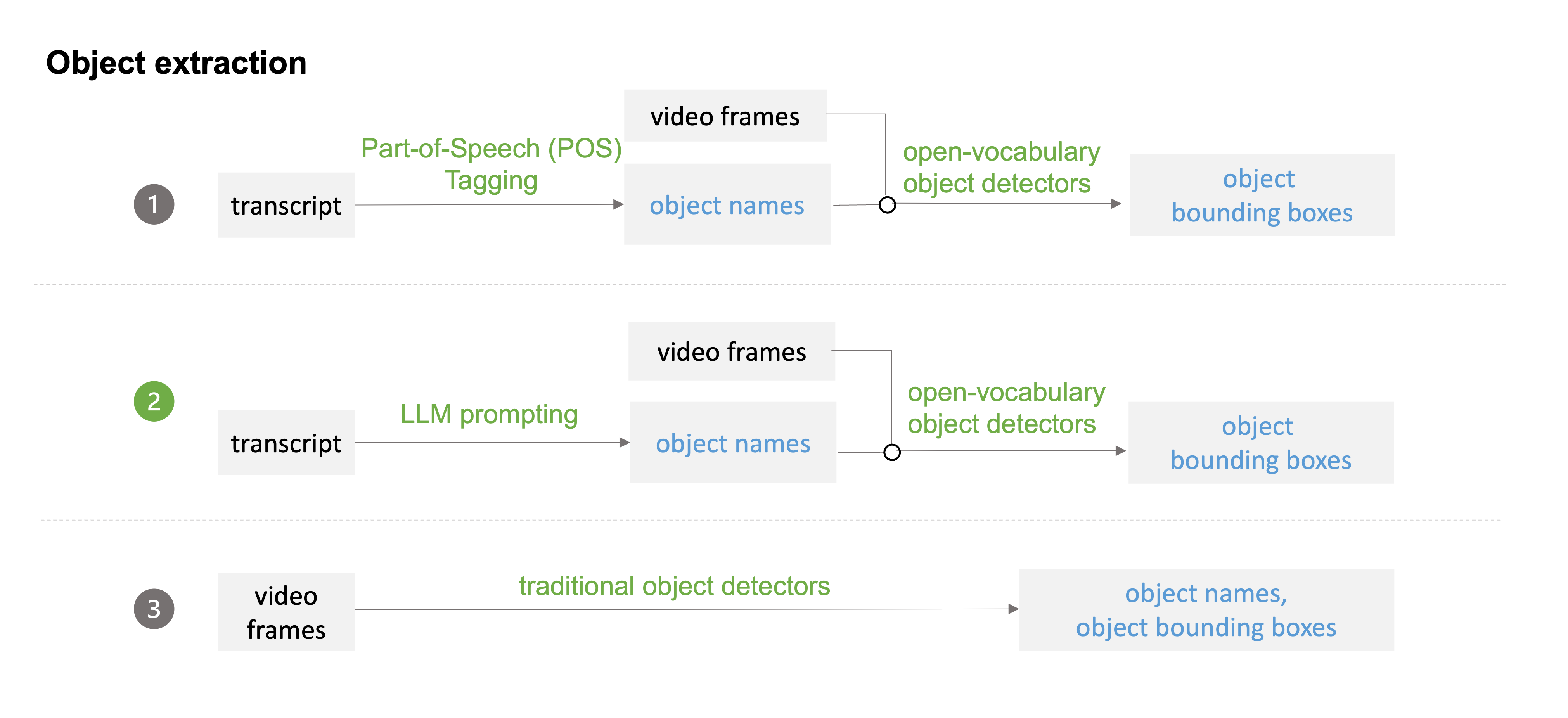}
    \caption{Three candidate pipelines for object extraction. Models are in green, and generated subcomponents are in blue. After evaluation, the chosen one is No.2.}
    \label{fig:object-extraction}
\end{figure*}

After considering the relevant models, we assemble them into three candidate pipelines.

\begin{itemize}
    \item \bpstart{Pipeline 1: POS Taggers + Open-vocabulary detectors}
As shown in Figure~\ref{fig:object-extraction}, pipeline (1) uses POS taggers to identify object names from the video transcript. It then passes these names \DIFadd{and} video frames into open-vocabulary object detectors to localize the objects.

  \item \bpstart{Pipeline 2: LLM + Open-vocabulary detector}
Pipeline (2) prompts an LLM to extract object names from the transcript and runs an open-vocabulary object detector.

  \item \bpstart{Pipeline 3: traditional object detectors} \DIFadd{Pipeline} (3) only uses traditional object detectors to obtain both the object names and bounding boxes.
\end{itemize}

\subsection{Evaluation of applicable models and candidate pipelines}

\subsubsection{Overall evaluation approach and metrics}

We evaluate models within the mentioned pipelines and discard any with subpar performance.  \DIFdel{We}\DIFadd{Based on available source code and pre-trained models, we} use at least one state-of-the-art (SoTA) implementation for each model type. While objective metrics are utilized, we also conduct manual inspections, especially when standard metrics fail to capture the error profiles. \DIFadd{In the following subsections, we report the main findings from the evaluation. Appendix A.1 provides detailed information about the evaluation dataset and results.}




\subsubsection{Evaluation dataset} 
 \DIFdel{Our evaluation dataset is }  \DIFadd{We evaluated on the validation set of} YouCook2~\cite{zhou2018towards},  \DIFdel{comprising 2000 untrimmed }  \DIFadd{containing 347} cooking videos with  auto-generated \DIFadd{English transcripts. Each video has human-annotated objects, step descriptions, and start/end times.} 


\subsubsection{Step pipeline evaluation} \hfill

\bpstart{Text descriptions: transcript summarization}
Pipeline 1, 2, and 3 rely on text summarization to derive step text descriptions. We assessed five methods, spanning both extractive (pulling key sentences from the source text\DIFadd{, e.g., LexRank~\mbox{
\cite{mihalcea2004textrank}}
, TextRank~\mbox{
\cite{erkan2004lexrank} }
}) and abstractive (rephrasing the original content, \DIFadd{e.g.,} BART~\cite{lewis2019bart}, T5~\cite{raffel2020exploring}, GPT-3~\cite{brown2020language}) \DIFadd{methods}. \DIFadd{Among the five methods,} GPT-3 leads by a large margin  \DIFadd{in ROUGE scores~\mbox{
\cite{lin2004rouge} }
(Appendix Table 5)}. 

Traditional NLP metrics might not effectively gauge the quality of text generated by LLMs~\cite{liu2023gpteval}. Through manual comparisons between \DIFadd{GPT-generated} descriptions and human annotations, we noted discrepancies that could affect ROUGE scores without necessarily compromising summarization quality. For instance:

\begin{itemize}
    \item LLM identifies optional steps, e.g., put the salad in the fridge.
    \item LLM turns states into steps, e.g., from the statement  
  \quotes{I've preheated my oven to 375 degrees}, it derived  \DIFadd{a step} \quotes{Preheat oven to 375 degrees}\DIFdel{and }
 .
    \item LLM includes more cooking details, e.g., temperature. 
\end{itemize}

Given this, we decided to select LLM for text summarization.

\bpstart{Text descriptions: video dense captioning} Pipeline 4 relies on dense captioning to obtain text descriptions. We evaluated two video dense captioning methods: MT~\cite{zhou2018end} and PDVC~\cite{wang2021end} \DIFadd{and there are evident errors} in object names and actions. For example, in the video "How to Make Fried Calamari | Hilah Cooking"\footnote{https://www.youtube.com/watch?v=-k7trpuj3X8}
\addtocounter{footnote}{-1},
the human annotation is ``drop the squid pieces into the oil'', but the dense captioning returns ``add the chicken in a pot of water boil''. Consequently, we decided not to incorporate dense captioning models, leading to the removal of pipeline 4.

\bpstart{Step timestamps} In the remaining pipelines, we evaluated models to identify timestamps: 
\DIFadd{NLVL method} DORi~\cite{rodriguez2021dori} (Pipeline 1) 
  \DIFadd{,} LLM prompting (GPT-3~\cite{brown2020language}) (Pipeline 2)  
\DIFdel{Shot boundary detectors: }  \DIFadd{and shot boundary detector} ProcNets~\cite{zhou2018towards} (Pipeline 3) 
  \DIFadd{.
}


 For pipeline 1, we provided the video and ground truth step descriptions to DORi~\cite{rodriguez2021dori} to predict \DIFadd{each step's} start and end time\DIFdel{of each step. The intersection over union (tIOU) for the generated segments compared to ground truth is 0.30 (a score closer to 1 implies a closer match). }\DIFadd{.} After manual inspection, we found that\DIFdel{it}  \DIFadd{the returned steps} did not observe the order (e.g., step 3 is localized before step 2) and returned overlapping steps. Given the considerable editing effort required for such errors, and \DIFdel{after reviewing literature that } other NLVL models \DIFdel{similarly neglect step order}\DIFadd{suffer from similar limitations}, we eliminated Pipeline 1.

For pipeline 2, we applied LLM alone to predict the boundary timestamps. We sent a transcript and a prompt ``\textit{summarize the video transcripts in several steps and find the start and end time for each step}''. The transcript format is the same as the YouTube transcript, with each sentence beginning with a timestamp. Since this approach predicts both the step summaries and timestamps simultaneously, complicating quantitative evaluation without timestamping all 347 videos manually. We sampled 20 videos and conducted a qualitative evaluation, showing LLM\DIFdel{did not return misordered and overlapping} \DIFadd{returns ordered and non-overlapping} steps, and the step descriptions and timestamps were reasonably matched with the ground truth.

For Pipeline 3, we employed ProcNets~\cite{zhou2018towards} to determine video shot boundaries. Relying solely on frame visuals, ProcNets scores each segment. We evaluated top-scored segments against the ground truth by computing the average temporal intersection over union (tIOU), however, given a low alignment (tIOU = 0.18), we didn't proceed to generate text summarization for each step.

\DIFadd{Therefore}, we retained Pipeline 2 for extracting steps.

\subsubsection{Object pipeline evaluation}
As shown in Figure~\ref{fig:object-extraction}, individual model types include POS taggers (pipeline 1), LLM prompting (pipeline 2), open-vocabulary detectors (pipeline 1 and 2) and traditional object detectors (pipeline 3).

\bpstart{Object names}
In Pipeline 1, we applied POS tagger Flair~\cite{akbik2019flair}  \DIFdel{and designated words labeled as NN, NNS, NNP, or NNPS~\mbox{
\cite{pennTreebank} }
as objects}  \DIFadd{to extract object names}. For Pipeline 2, we prompted GPT-3~\cite{brown2020language, ouyang2022training} with the transcript and an instruction: \quotes{Identify the objects, ingredients, tools, equipment in this tutorial} and parsed objects from the response. In Pipeline 3, we \DIFdel{down-sampled videos to one frame every 10 seconds and}employed a faster R-CNN~\cite{ren2015faster} trained on the Visual Genome dataset~\cite{krishna2017visual}. Both POS taggers and GPT-3  \DIFdel{identified at least 7 objects, outperforming visual detectors }  \DIFadd{outperformed visual detectors in identifying true positives}. However,  \DIFdel{both POS taggers and visual detectors }  \DIFadd{POS taggers} often identified non-cooking objects, \DIFdel{producing 43.6 and 32.4 irrelevant objects (} e.g., the chef's necklace  \DIFdel{), respectively, and GPT-3 only had 6.8}  \DIFadd{(Appendix Table 6)}. As such, we retained only Pipeline 2, leveraging LLM for object extraction.

\DIFdelFL{Method }
\DIFdelFL{True positives }
\DIFdelFL{Label unavailable }
\DIFdelFL{Missing }
\DIFdelFL{False positives  }
\DIFdelFL{Visual Detector~\mbox{
\cite{ren2015faster} }
}
\DIFdelFL{2.8 }
\DIFdelFL{2.9 }
\DIFdelFL{4.2 }
\DIFdelFL{43.6 }
\DIFdelFL{POS taggers~\mbox{
\cite{akbik2019flair} }
}
\DIFdelFL{7.0 }
\textbf{\DIFdelFL{1.1}} 
\textbf{\DIFdelFL{1.5}} 
\DIFdelFL{32.4 }
\DIFdelFL{GPT-3 with prompt }
\textbf{\DIFdelFL{7.4}} 
\textbf{\DIFdelFL{1.1}} 
\textbf{\DIFdelFL{1.5}} 
\textbf{\DIFdelFL{6.8}}
{
\DIFdelFL{A quantitative comparison of object detection methods. Label unavailable: the object is not in the Visual Genome~\mbox{
\cite{krishna2017visual} }
dataset or is unmentioned in the transcript. Missing: fails to detect the object when the label is available. False positives: detections irrelevant to the cooking process.}}

 \bpstart{Object bounding boxes} 
Considering the underwhelming results of traditional object detectors, we only evaluated open-vocabulary object detectors \DIFadd{and eventually chose} OWL-ViT~\cite{minderer2022simple} \DIFadd{considering both performance and computational cost.}

\begin{figure*}
    \centering
    \includegraphics[width=\textwidth]{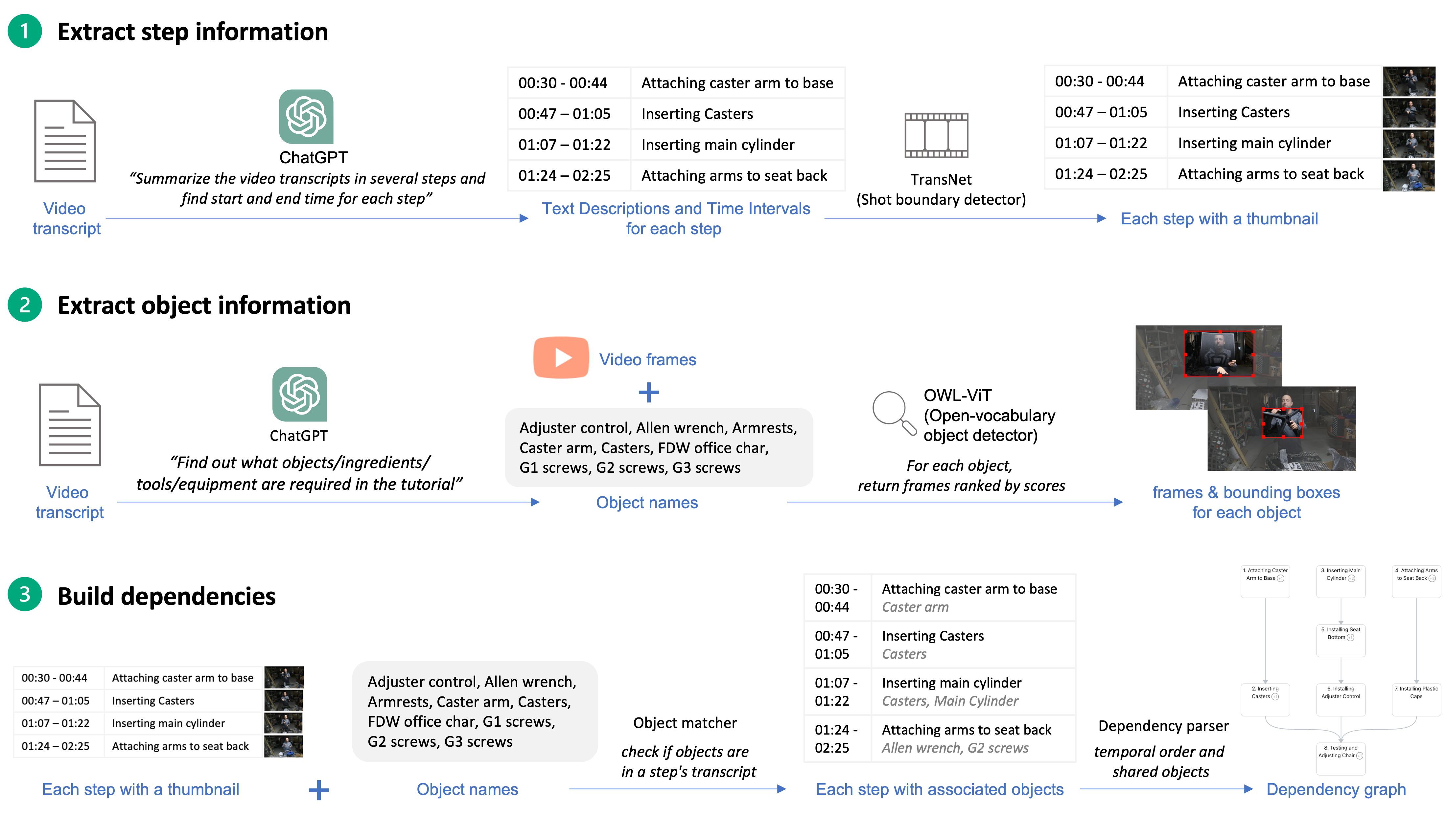}
     \caption{TutoAI's machine learning pipelines to obtain objects and steps in instructional videos: 1. extract steps: ChatGPT processes the video transcript to produce text descriptions and time intervals for each step, then a shot boundary detector augments each step with a thumbnail; 2. extract objects: ChatGPT identifies the objects in the tutorial, then an open-vocabulary object detector returns the frames and bounding boxes of the objects; 3. build dependencies: an object matcher checks if objects are in a step's transcript and produces a dependency graph.}
     \label{fig:mlDiagram}
  \end{figure*}

\subsection{Final pipeline}
 \DIFdel{Based on the above results, we }  \DIFadd{We} finalized our pipeline as shown in Figure~\ref{fig:mlDiagram}, which includes Step pipeline 2 (Figure~\ref{fig:step-extraction}) and Object pipeline 2 \DIFdel{in Figure~\mbox{\ref{fig:step-extraction}}}  \DIFadd{( Figure~\mbox{\ref{fig:object-extraction})}} . First, we extract \DIFadd{steps} from video transcripts by prompting LLM (here we use \DIFadd{GPT-3.5} ~\cite{chatGPT}, assuming it has better performance than GPT-3): ``Summarize the video transcripts in several steps and find start and end time for each step,'' then we use a shot boundary detector~\cite{souvcek2020transnet} to pick thumbnails for each step. Next, to extract object components, we make a different prompt: ``Find out what objects/ingredients/ tools/ equipment are required in this tutorial.'' Then, we run an open-vocabulary detector~\DIFadd{\mbox{
\cite{huggingfaceowlvit}}} to identify the bounding boxes in video frames. Finally, we match object names to each step's description via string match, then build dependencies between steps by the shared objects.

%% file: 6-layer-3-UI.tex
\section{Level 3: User Interfaces for Mixed-Media Tutorial Creation}
\label{sec:ui}

 \subsection{\DIFadd{Design considerations}}
 \label{sec:UI-design-considerations}
 Section 4 shows various mixed-media tutorial formats regarding visual representation, layout, and interactivity tailored to specific domains. Rather than advocating a one-size-fits-all format, we embrace the principle of \textit{separating content from style}: mixed-media tutorial components are content that can be extracted, reviewed, and edited, with different styles (e.g., visual representations, layouts, and interactive behaviors) added later. We focus on enabling creators to inspect and modify content, assuming that a tool will auto-apply styles to the final tutorial. Thus, we propose the following UI design considerations to elevate the creator experience without information overload (\textbf{D3}).

\begin{itemize}
    \item[C1] \textbf{Component-based creation. } The UI should break down the creation process into individual tasks based on the  \DIFdel{components in the mixed media tutorial }  \DIFadd{mixed-media tutorial components}. The UI should sequence tasks  \DIFdel{such} \DIFadd{so} that the output from one task \DIFdel{(e.g., step extraction) }can provide context to help users perform subsequent tasks\DIFdel{(e.g., dependency building) more} efficiently.

    \item[C2] \textbf{One modality at a time.}
    To reduce context switching, 
    when a component encompasses multiple modalities (i.e., text and images), the UI should break it down into subtasks. This will help simplify user interactions and avoid requiring users to operate across multiple modalities in a single task.

    \item[C3] \textbf{Editable AI output.} The UI should enable creators to keep, modify, or dismiss AI-generated results and add information missed by AI.
    \item[C4] \textbf{Real-time edit preview.} Upon editing, the UI should automatically reflect changes in the tutorial.
\end{itemize}


 \subsection{\DIFadd{An example prototype}}
\begin{figure*}[htbp]
    \centering
    \includegraphics[width=.95\textwidth]{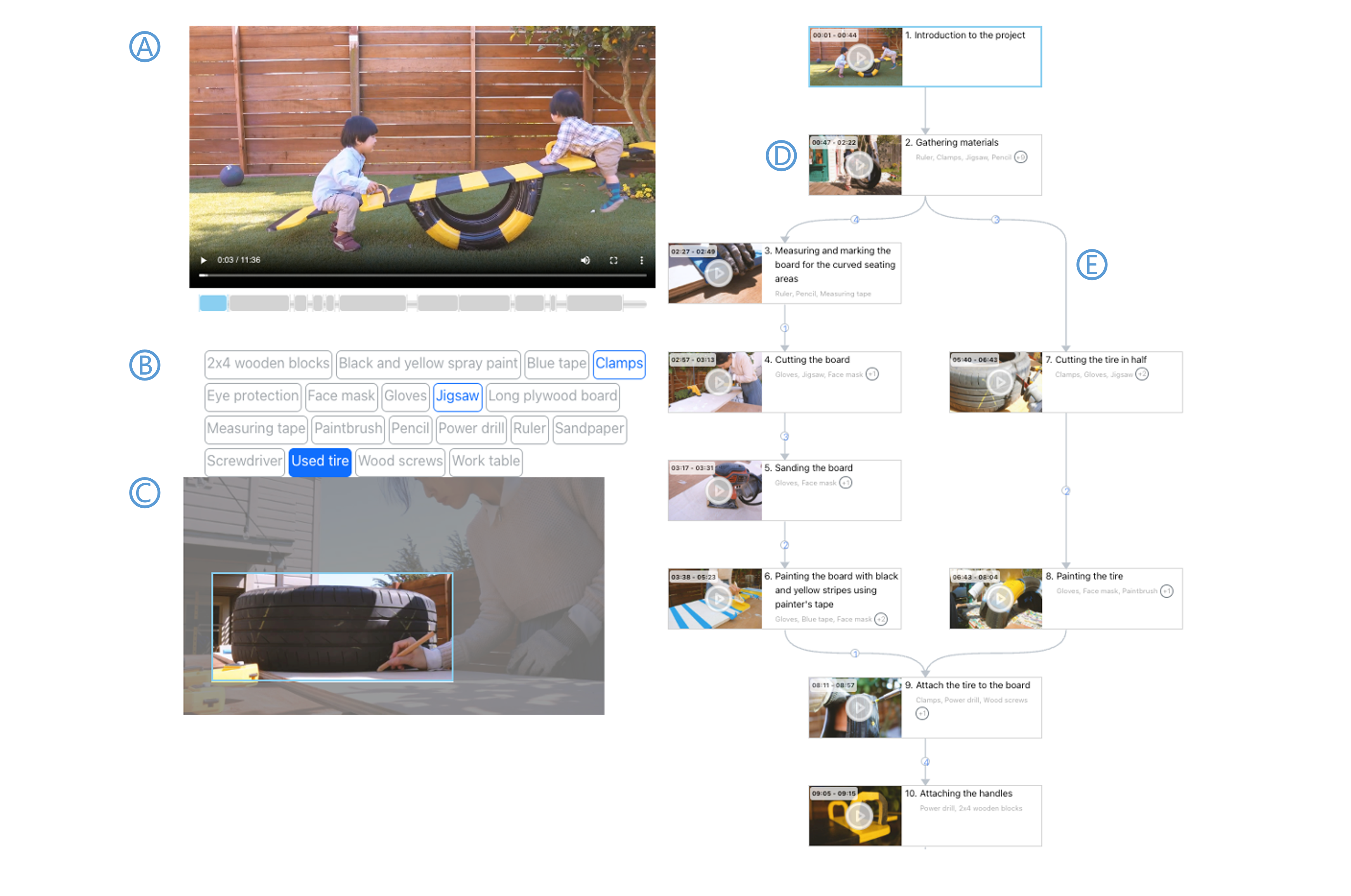}
    \caption{A mixed-media tutorial template on making a seesaw for kids: below the video player (A) is a list of required objects (B); hovering on the blue-bordered object will show the object's image along with a bounding box (C); 
    on the right is an overview of steps, (D) each step is a video clip with start and end time, text descriptions and associated objects. (E) The arrows between the steps indicate the dependencies. }
    \label{fig:seesaw}
\end{figure*}

 We reify these design guidelines into an example UI  \DIFdel{for mixed media tutorial creation of physical tasks } and use the video ``How to make a seesaw for kids''\footnote{https://www.youtube.com/watch?v=drDSY3ZZqnQ, used with permission} as  \DIFdel{our driving } input. In this implementation, we use a tutorial format depicted in Figure~\ref {fig:seesaw}. The tutorial contains the following components: a video player and step boundary below it (Figure~\ref {fig:seesaw}A)\DIFadd{, }an object list (Figure~\ref {fig:seesaw}B) over which users can hover to see an image of the selected objects (Figure~\ref {fig:seesaw}C); step overviews, which consist of a text description, a representative thumbnail and objects for each step (Figure~\ref {fig:seesaw}D); associated dependencies (Figure~\ref {fig:seesaw}E), represented as arrows between steps, and the buttons on the arrow show objects that connect steps. We chose this tutorial design for its comprehensive components without domain-specific assumptions.

\begin{figure*}
    \centering
    \includegraphics[width=\textwidth]{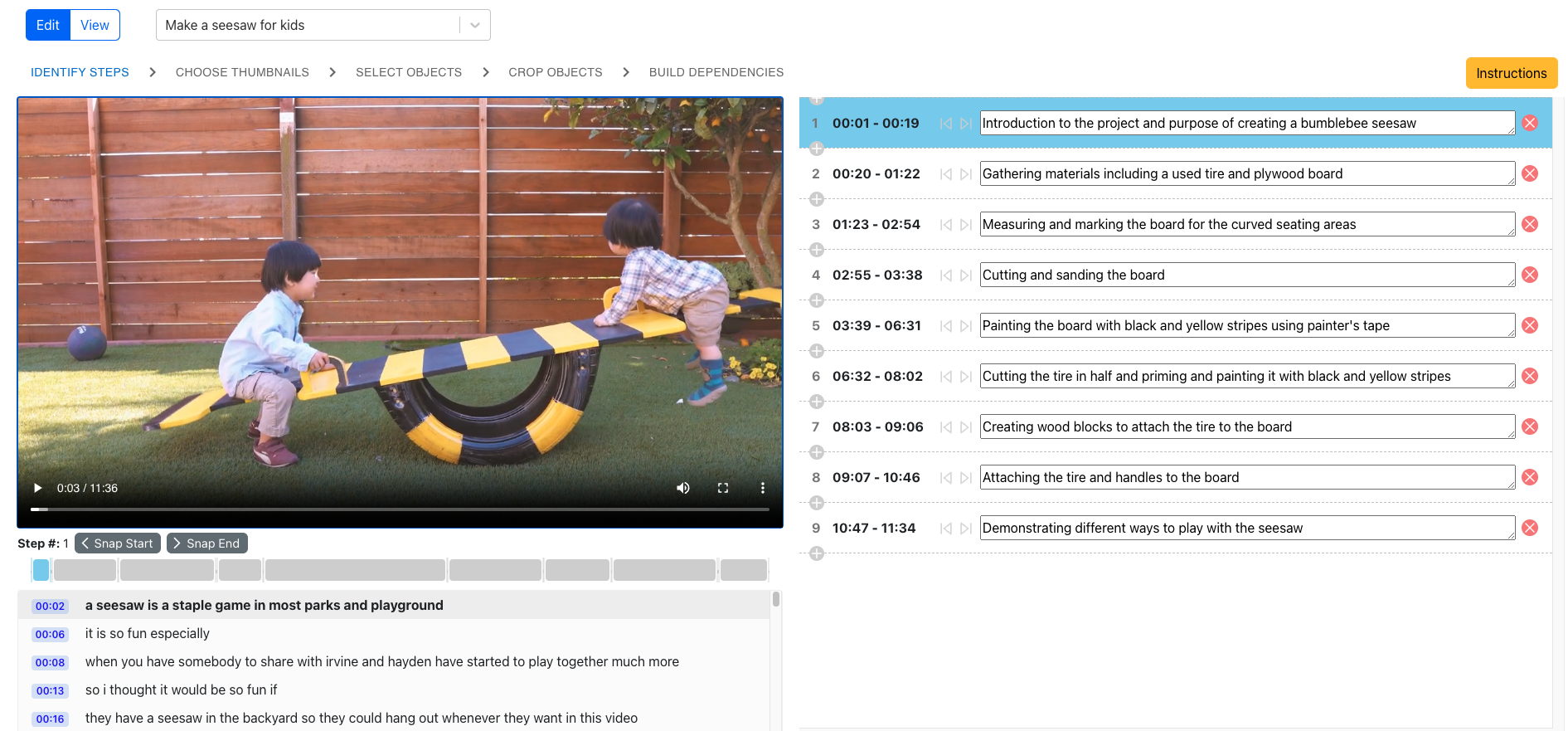}
    \caption{Identify steps. This task aims to break down the video into several steps and provide text descriptions and time boundaries for each step. On the left is a video player and its transcript (“Make a seesaw for
kids”); on the right are the AI-generated steps.}
    ~\mbox{\label{fig:stage1}}
\end{figure*}

 \DIFdel{For structured guidance~}\textbf{\DIFdel{(C1)}}
\DIFdel{, the }  \DIFadd{The} UI breaks up the creation process into five sequential tasks, each targeting a single tutorial component -- steps, objects, or dependencies -- in a single modality~\textbf{(C2)}. 
\DIFdel{Here is a brief description of the workflow in our UI }  \DIFadd{Creators can bypass any tasks and accept the default results if they deem the task unnecessary }\textbf{\DIFadd{(C3)}}\DIFadd{. As they make changes, creators can preview the updates with the current modifications ~}\textbf{\DIFadd{(C4)}} \DIFadd{by the "view" button (Figure~\mbox{\ref{fig:stage1}}). Here is the workflow}\DIFdel{.
}\DIFadd{:} 

 \bpstart{1) Identify steps}  The UI shows the video and its transcript on the left, AI-generated steps with text descriptions and start/end timestamps on the right (Figure~\ref{fig:stage1}); creators can edit the text, add/delete steps, and update the time boundaries by dragging the range slider~\textbf{(C3)}.  

\bpstart{2) Choose step thumbnails}
The UI presents dissimilar candidate video frames. 
Creators can adjust the number of frames using a ``show more/less''slider, and select a frame. (Appendix Figure 12). The thumbnails presented for a given step are bounded by the time boundaries identified for that step in task 1 ~\textbf{(C1)}.

\bpstart{3) Select objects}
The UI suggests an object list required for the tutorial and associates the objects with the steps (Appendix Figure 13). Creators can modify objects and change their step associations~\textbf{(C3)}.

\bpstart{4) Crop objects} Creators can choose a representative image for each object (Appendix Figure 14). The UI shows a list of objects refined by users in task 3 ~\textbf{(C1)} and presents candidate frames with probable object bounding boxes, which creators can adjust ~\textbf{(C3)}. 

\bpstart{5) Build dependencies}
The final task is to build dependencies (Appendix Figure 15). The UI displays a node-link diagram of dependencies based on shared objects between the steps, as identified in task 3 ~\textbf{(C1)}. Creators can add/delete links via drag and drop ~\textbf{(C3)}.  

 \DIFdel{Creators can bypass any tasks and accept the default results if they deem the task unnecessary ~}\textbf{\DIFdel{(C3)}}
\DIFdel{. As they make changes, creators can preview the updates with the current modifications ~}\textbf{\DIFdel{(C4)}} 
\DIFdel{by the "view " button (Figure~\mbox{\ref{fig:stage1})}}  


%% file: 7-framework-evaluation.tex
\section{TutoAI framework evaluation: Model} \label{sec:model}

To demonstrate our pipeline's generality, we evaluated it on a small yet diverse dataset.

\subsection{Dataset}
Inspired by the object-action quadrant for instructional videos~\cite{chang2021rubyslippers}, we considered the following diversity dimension of instructional videos: \DIFadd{creator, task}, video duration, number of steps, number of objects\DIFdel{narration style (instructional or conversational)}. \DIFadd{The content creator dimension allows us to capture variations over editing styles such as instructional or conversational narration, concise versus verbose steps, use of music fillers, etc.}
As a result, we collected a dataset of \DIFdel{8} \DIFadd{20} 
\DIFadd{videos} (Table~\ref{tab:groundTruthComparison}) across four domains: cooking, crafting, makeup, and repair. \DIFadd{Each video within a domain focused on a different task (e.g., fixing an iPhone vs. fixing a hole in the wall for repairs) and was made by a different creator.}
We manually annotated the 1) objects and 2) step boundary timestamps and used these as ground truths. 
We assessed our pipeline on object extraction and timestamp prediction.

\subsection{Object extraction results}

We compare object extraction results with the ground truth using the F1 score, computed as:
\[ F1(o_{ours}, o_{gt}) = \frac{2 |o_{ours}\cap o_{gt}|}{|o_{ours}|+|o_{gt}| }\]
where $o_{ours}$ is the set predicted by our pipeline and $o_{gt}$ is the ground truth, and $|o|$ denotes the number of objects in the set. 
As shown in Table~\ref{tab:groundTruthComparison} column 8 (``F1''), our object extraction F1 scores fall between 0.56 to 1, with an average of \DIFdel{0.80}\DIFadd{0.88}, indicating great performance across domains. \DIFadd{False negatives often resulted from objects not explicitly referenced in the transcript.}

\subsection{Step boundaries} 

Our pipeline outputs a sequence of steps, including text descriptions and start and end timestamps. On average, it \DIFdel{has}\DIFadd{yields} \DIFdel{1.37}\DIFadd{1.3} false negative steps \DIFadd{and 0.25 false positive steps} per video (Table~\ref{tab:groundTruthComparison} column 11 ``\# False Neg.'' \DIFadd{and column 12 ``\# False Pos.''}). \DIFdel{Introduction or conclusion segments account for $54.5\%$ of the false negatives, while the remaining $45.5\%$ are true steps. No false positive step descriptions were generated.} \DIFadd{The low false negative and false positive rates suggest that our pipeline does a good job of extracting steps. Introduction and conclusion segments accounted for most false negative steps, and false positive steps were incorrectly inferred from verbose narrations.} We then used F1 score to assess predicted timestamps against the ground truth. For false negative steps, we set $t_{ours}$ to $[0,0]$ to signify that this step did not appear. Aggregate F1 scores ranged from \DIFdel{0.39 to 0.80, averaging 0.65}\DIFadd{0.22 to 0.95, averaging 0.59} (Table~\ref{tab:groundTruthComparison} column \DIFdel{12}\DIFadd{13} ``Avg. F1'') \DIFdel{suggesting good alignment with predicted step boundaries and ground truth}. \DIFadd{In general, we found that our pipeline performed better on the step localization task for shorter tutorials and tutorials with more concise steps. Certain video editing decisions, such as using non-speech fillers between steps, showing step execution before verbally describing it, and describing steps out of order, also negatively impacted localization. Our aggregate F1 score suggests reasonable alignment between predicted step boundaries and ground truth with room for improvement, which can be achieved via more sophisticated prompt engineering.}   



\begin{table*}[htbp]
    \centering
    \begin{tabular}{lcc|ccccc|ccccc}
    \toprule
    & & & \multicolumn{5}{c}{\textbf{Objects}} & \multicolumn{5}{c}{\textbf{Steps}} \\
    & & Duration & Ours & GT & False & False & & Ours & GT & \# False & \DIFadd{False} & Avg. \\
     Video ID & Domain & (minutes) & \# Obj. & \# Obj. & Neg. & Pos. & F1 & \# Steps & \# Steps & Neg. & \DIFadd{Pos.} & F1\\
     \hline
     \DIFadd{36FOyZ26ld0} & cooking & 0:24 & 10 & 10 & 0 & 0 & 1 & 4 & 5 & 1 & 0 & 0.95\\
     j4UVB6MPsKw & cooking & 5:27 & 16 & 16 & 3 & 3 & 0.81 & 6 & 6 & 0 & 0 & 0.80\\
     BAp1AXn82Pg & cooking & 7:32 & 20 & 23 & 3 & 0 & 0.93 & 8 & 9 & 1 & 0 & 0.72 \\
     \DIFadd{Y-Y9CXGRJPU} & cooking & 13:50 & 24 & 26 & 3 & 1 & 0.92 & 9 & 12  & 3 & 3 & 0.34\\
     \DIFadd{L0Gu2KDCS6o} & cooking & 15:10 & 17 & 19 & 2 & 0 & 0.94 & 9 & 12 & 3 & 0 & 0.22\\
     \DIFadd{zQ8gThfBDqU} & crafting & 3:40 & 12 & 14 & 2 & 0 & 0.92 & 11 & 11 & 0 & 0 & 0.69\\
     OUMfV1D0\_RQ & crafting & 4:58 & 8 & 6 & 1 & 3 & 0.71 & 9 & 9 & 0 & 0 & 0.72\\     
     \DIFadd{SX4DCFDKMzc} & crafting & 7:48 & 13 & 18 & 6 & 1 & 0.77 & 13 & 13 & 0 & 0 & 0.65\\
     DU4DiGeLr6Y & crafting & 10:21 & 5 & 5 & 0 & 0 & 1 & 6 & 7 & 1 & 0 & 0.74\\
     \DIFadd{VKZI7X-UIe8} & crafting & 18:55 & 17 & 19 & 2 & 0 & 0.94 & 7 & 8 & 1 & 0 & 0.52\\
     Ls969BmW1kw & makeup & 5:00 & 13 & 13 & 3 & 3 & 0.77 & 9 & 12 & 3 & 0 & 0.57\\
     \DIFadd{skZ-nUB\_b00} & makeup & 5:26 & 10 & 12 & 2 & 0 & 0.91 & 13 & 13 & 0 & 0 & 0.70\\
     QmPiBCu5\_ME & makeup & 7:49 & 16 & 18 & 2 & 0 & 0.94 & 10 & 12 & 2 & 0 & 0.71\\
     \DIFadd{gkkmHizG2As} & makeup & 13:10 & 8 & 9 & 1 & 0 & 0.94 & 6 & 6 & 0 & 0 & 0.69\\
     \DIFadd{9f7zmCSzG9E} & makeup & 13:26 & 25 & 25 & 2 & 2 & 0.92 & 8 & 11 & 3 & 0 & 0.42\\
     \DIFadd{lj7YK1lIRUM} & repair & 2:23 & 16 & 16 & 0 & 0 & 1 & 14 & 15 & 1 & 0 & 0.81\\
     ZWlq\_fWRrzI & repair & 4:09 & 9 & 7 & 1 & 3 & 0.75 & 7 & 9 & 2 & 0 & 0.39\\
     B4iWwUzxFWA & repair & 4:17 & 5 & 13 & 8 & 0 & 0.56 & 4 & 6 & 2 & 0 & 0.61\\
     \DIFadd{p55lnFCorQ4} & repair & 9:57 & 11 & 9 & 1 & 3 & 0.8 & 12 & 15  & 3 & 2 & 0.31\\
     \DIFadd{b-GLI-Vsu9s} & repair & 11:38 & 11 & 12 & 1 & 0 & 0.96 & 10 & 10 & 0 & 0 & 0.33\\
      
     \bottomrule
    \end{tabular}
    \caption{Pipeline evaluation on ground truth. We annotate ground truth for 20 instructional videos from 4 different domains and test the object extraction and step boundary detection components of our pipeline on these videos. Our pipeline performs object extraction very well (average F1 = 0.88) across domains. Our steps boundary detection performs relatively well on at least one video in each domain (F1 = 0.59).}
    \label{tab:groundTruthComparison}
\end{table*}

%% file: 8-framework-evaluation.tex
\section{TutoAI framework evaluation - UI}
To evaluate the quality of AI-extracted components perceived by users and the tutorial creation experience, we conducted two preliminary user studies to understand 1) if the TutoAI framework generates higher-quality mixed-media tutorial components than a baseline method before editing, 2) if the TutoAI framework generates mixed-media tutorials that are more useful for consumers than a baseline method after editing, and 3) the potential of integrating TutoAI into creators' existing workflow. 



\subsection{\DIFadd{Study design rationales}}
We identify both instructional video consumers and influencers who make instructional videos as potential users of our prototype. Video consumers who want to learn instructional content
are motivated to interact with the mixed-media tutorials and can benefit from tutorial creation. For example, Kim et al. find that when students contributed to creating subgoal-based tutorials, they became more attentive to learning~\cite{weir2015learnersourcing}; popular video platforms also support video consumers to create video clips (e.g., YouTube's ``create clip''\footnote{https://support.google.com/youtube/answer/10332730}) and mixed-media notes (e.g., Coursera's ``save note''\footnote{https://blog.coursera.org/ready-for-retention-presenting-a-unified-note-taking-experience/}). 
Therefore, we recruited participants who frequently watch instructional videos for study 1. Several participants also disclosed that they had created mixed-media tutorials before, confirming our assumption. For study 2, we recruited two YouTube creators who regularly publish instructional videos. 

In both studies, we used auto-generated YouTube Chapters~\cite{vchapters} as the baseline. Although TutoAI was inspired by previous works, these tutorials were either generated automatically using a domain-specific approach~\cite{chi2012mixt, wang2014evertutor, fraser2020temporal, truong2021automatic, kim2014data} or manually without AI assistance~\cite{liu2018conceptscape, yang2022improving}. Mixed-initiative approaches~\cite{pavel2014video, kim2014crowdsourcing, chang2018recipescape, nawhal2019videowhiz} do not provide comparable creation experience like TutoAI. We thus determined that YouTube Chapters~\cite{vchapters} is the most reasonable baseline since they also support cross-domain generation of steps.

\subsection{Study 1: general users}

\subsubsection{Recruitment:} we recruited 24 participants (female: 10, male: 13, non-binary: 1) who regularly watch instructional videos on YouTube (several times a week: 7, several times a month: 14, several times a year: 3). They watch instructional videos in various domains: cooking (19), home projects (15), software \& programming (15), sports \& fitness (13),  electronics (9), beauty (6), and animals \& pets (3). 12 participants have used the YouTube Chapter feature. \DIFadd{Though not prolific YouTube creators, five participants have created video tutorials: for a mobile app (P1), cooking (P5), robots (P11), design tools (P19), and Android development (P20)}. 

\subsubsection{Instructional videos:} we chose two instructional videos on YouTube: office chair assembly\footnote{https://youtu.be/OEIDupReh8Q} and strawberry blueberry shortcakes\footnote{https://youtu.be/BAp1AXn82Pg}. We randomly split the participants into two groups: A (office chair assembly, video length: 5 minutes 18 seconds) and B (strawberry blueberry shortcakes, video length: 7 minutes 32 seconds). Participants' median familiarity with the video topic was 2.5 and 3.0, respectively (1: not familiar at all, 5: extremely familiar). 

\subsubsection{Procedures:} First, we briefly introduced the concept of mixed-media tutorials and editing features of the UI, 
then, participants followed a step-by-step instruction to reproduce a Kung Pao chicken\footnote{from the YouCook2 dataset: https://youtu.be/ntiGX3X-spA} mixed-media tutorial created by TutoAI as a warm-up. Then, the participants were asked to create a mixed-media tutorial for the assigned video and think aloud. Next, participants completed a survey and provided open-ended feedback. Each session was remotely conducted over Zoom and lasted about 1 hour. Each participant received a \$20 Amazon gift card. The study was approved by the Institutional Review Board (IRB) Committee.

\subsubsection{Findings:} 
We observed that participants applied different strategies to create mixed-media tutorials. Some participants watched the entire video first, some watched each step's video clip based on the AI-generated results first, and some did not watch the video but read the transcript instead. 

\begin{figure}
  \centering
  \begin{subfigure}{0.4\textwidth}
    \centering
    \includegraphics[width=\textwidth]{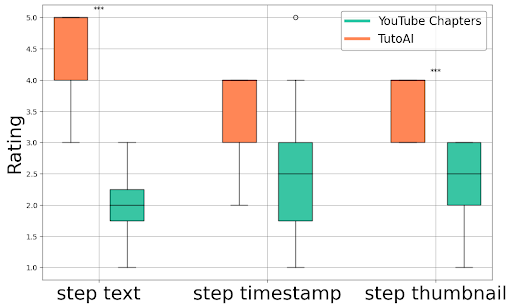}
    \caption{Quality before editing, TutoAI vs. YouTube Chapters, text: 4.6$\pm$0.65 vs. 2.0$\pm$0.71 ($p$=0.009); timestamps: 3.5$\pm$0.65 vs. 2.5$\pm$1.19 ($p$=0.075); thumbnails: 3.6$\pm$0.49 vs. 2.3$\pm$0.75 ($p$=0.021)} 
    \label{fig:grp-a}
  \end{subfigure}%
  \hspace{5mm}
  \begin{subfigure}{0.4\textwidth}
    \centering
    \includegraphics[width=\textwidth]{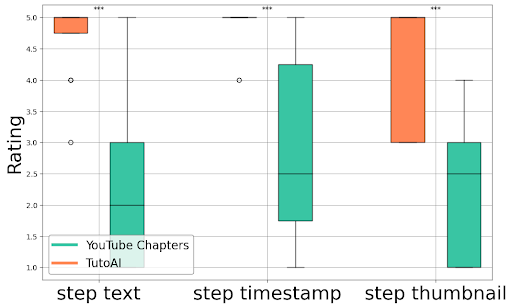}
    \caption{Usefulness after editing, TutoAI vs. YouTube Chapters, text: 4.7$\pm$0.62 vs. 2.3$\pm$1.25 ($p$=0.003), timestamps: 4.9$\pm$0.28 vs. 2.8$\pm$1.52 ($p$=0.021), thumbnails: 4.3$\pm$0.94 vs. 2.2$\pm$1.16 ($p$=0.015)}
    \label{fig:grp-b}
  \end{subfigure}
  \caption{Component quality of group A: office chair assembly. Before editing (left), after editing (right).}
  \label{fig:before-quality}
\end{figure}

\bpstart{Quality of AI-generated results} 
We asked the participants to rate the quality of components generated by TutoAI \textit{before editing} and YouTube auto-generated Chapters on a five-point Likert scale, 
where 1 means ``the quality is so low that the author needs to start from scratch'', and 5 means ``the quality is so high that the author barely needs to do anything''. YouTube Chapters only generates timestamps, thumbnails, and text descriptions for each step. \DIFadd{We conducted a Wilcoxon Signed-Rank test with a Bonferroni correction, and found TutoAI generated higher quality results than YouTube chapters in 2/3 comparisons in group A} (Figure~\ref{fig:grp-a}): TutoAI vs. YouTube Chapters, text: 4.6$\pm$0.65 vs. 2.0$\pm$0.71 ($p$=0.009); timestamps: 3.5$\pm$0.65 vs. 2.5$\pm$1.19 ($p$=0.075); thumbnails: 3.6$\pm$0.49 vs. 2.3$\pm$0.75 ($p$=0.021). For group B, the benefits of TutoAI are not statistically significant  (Appendix Figure 17 (a)): TutoAI vs. YouTube Chapters, text: 4.4$\pm$0.64 vs. 3.6$\pm$1.04 ($p$=0.138); timestamps: 3.3$\pm$1.25 vs. 3.0$\pm$1.0 ($p$=1.000); thumbnails: 3.4$\pm$0.76 vs. 2.4$\pm$1.38 ($p$=0.138). Other scores of TutoAI components are in Appendix Figure 18.


\bpstart{Perceived Usefulness of Tutorial Components} 
We asked participants to rate each component's usefulness for tutorial consumers \textit{after editing}, where 1 refers to ``I don't think consumers will benefit from this component,'' and 5 refers to ``I'm confident that consumers will benefit from this component.'' \DIFadd{We conducted a Wilcoxon Signed-Rank test with a Bonferroni correction, and found TutoAI results more useful than YouTube Chapters in 3/3 comparisons in group A (Figure~\ref{fig:grp-b})}. Specifically, TutoAI vs. YouTube Chapters, text: 4.7$\pm$0.62 vs. 2.3$\pm$1.25 ($p$=0.003), timestamps: 4.9$\pm$0.28 vs. 2.8$\pm$1.52 ($p$=0.021), thumbnails: 4.3$\pm$0.94 vs. 2.2$\pm$1.16 ($p$=0.015). For group B, the benefits of TutoAI are not statistically significant. TutoAI vs. YouTube Chapters, text: 4.8$\pm$0.43 vs. 3.8$\pm$1.16 ($p$=0.063), timestamps: 4.8$\pm$0.37 vs. 4.0$\pm$1.22 ($p$=0.192), thumbnails: 4.0$\pm$0.91 vs. 2.6$\pm$1.50 ($p$=0.153). Other scores of TutoAI components are in the Appendix Figure 19. 

\bpstart{TutoAI vs. YouTube Chapters}
\DIFadd{
Although TutoAI has received higher scores than YouTube Chapters in both videos in the user study, the statistical results are insignificant for the strawberry blueberry shortcake video. We looked into the user study recordings and found that since text descriptions of YouTube Chapters are very short (``Strawberry topping'' and ``Chantilly cream''), the participants deem them to be helpful as long as they contain important keywords.
In comparison, the step descriptions generated by TutoAI are ``Preparing the strawberries for the topping'' and ``Preparing the Chantilly cream using an air disc container''. Although TutoAI provided more details, the participants believe the essential keywords have been captured by YouTube Chapters.
On the other hand, the YouTube Chapters for the office chair assembly video missed most keywords, e.g., ``Base Assembly'', and were deemed less useful than TutoAI-generated text descriptions: ``Attaching Caster Arm to Base''. To more conclusively demonstrate the superiority of the fine-grained text descriptions generated by TutoAI, we need more experiment data involving more instructional videos.}

\bpstart{Dependencies and other components} Many participants (17/24) found the dependency diagram useful (rated 4 or 5), e.g., P12 said \textit{``The flow charts were amazing...if I didn't want to watch the video, I could just see the steps...I am getting a visual representation of the whole video.''} While some expressed confusion, P4 said \textit{``dependency diagram was a bit tricky to understand.''} Besides existing components, participants also brainstormed new tutorial components, e.g., 
3D object augmentation/more camera angles (P11). 

\bpstart{Application Scenarios}  The participants shared situations where they would like to have a mixed-media tutorial, e.g., build a pet snake vivarium (P5) and collaborative software development (P8). Some participants also mentioned situations where they would like to create a mixed-media tutorial to refresh their memory, e.g., P9 said \textit{``I make quilts, and I have to look up a lot of tutorials for how to finish the quilt because you only do it once every time.''}. 

\subsection{Study 2: YouTubers}
\subsubsection{Preparation:} we recruited two YouTube creators (E1 and E2) who regularly publish instructional videos. For each YouTuber, we picked several of their videos with auto-generated YouTube Chapters. We ran our ML pipeline on the video: ``bike rack installation''\footnote{https://youtu.be/5nHD0vy9R5g, used with permission} (E1) and ``how to make a seesaw for kids''\footnote{https://youtu.be/drDSY3ZZqnQ, used with permission} (E2) and loaded the results into TutoAI UI. During the study, we briefly introduced mixed-media tutorials and asked them to complete a step-by-step warm-up task to get familiar with the UI. Then, they created a mixed-media tutorial for the video and provided oral feedback along the way. Each participant received a \$50 Amazon gift card.

\subsubsection{Findings:} we asked them about the impression of AI-generated results and workflows in creating instructional videos.

\bpstart{TutoAI vs. YouTube auto-generated Chapters} Both YouTubers spoke highly of the TutoAI-generated results, e.g., when asked about the quality of steps, E1 said \textit{``I'd say probably about a 4 (out of 5). There were a few things I changed, but for the most part, it was a good starting point.''}. When shown the auto-generated YouTube Chapters, E1 gave them a 2.5 to 3: \textit{``the first few are getting the breaks pretty good, but they lost some of the steps that your software captured''}. E2 believed it needs a redo completely: \textit{``I won't be able to use any of this...``Wood blocks'' is just the name of the material, not something meaningful for the viewers to imagine''}. The author-created steps are in Appendix Figure 16.

\bpstart{Attitudes towards dependencies} E1 expressed enthusiasm in applying dependency diagrams: \textit{``I really like the dependency diagram, especially for a procedural how-to video...it helps them understand... when you might need to skip a step or there might be a branch...''}. E2 saw the dependency diagram has better use in cooking videos, \textit{``for example, cooking...you can do many things at the same time. But for my (DIY) tutorial, it kind of depends on one flow.''}  

\bpstart{Incorporate TutoAI into existing workflow} 
We asked both E1 and E2 to share their thoughts on incorporating TutoAI into their workflow. E1 said \textit{``I think this is a great tool... I don't know that it would necessarily save me time just creating chapters. It's a different animal because this is giving me the ability to do a lot more, especially creating the flow charts, which I really like... viewers would get a lot out of this as opposed to just a regular chapter''}. E2 recounted that in the past, she spent about 1 hour writing down steps and time boundaries of a 10-min video she created (6 times of the original video length), and to her relief, with the help of TutoAI, it only took her 17.5-minutes to finalize steps and time boundaries for an 11.5-minute video (1.5 times of the original video length).

%% file: 9-discussion.tex
\section{Discussion}
 \DIFdel{TutoAI is the first framework trying to provide a unified approach to }  \DIFadd{We have proposed TutoAI, the first cross-domain framework for} AI-assisted mixed-media tutorial creation.  \DIFdel{Here, we discuss the implications of }  
\DIFadd{TutoAI extends earlier efforts in generalizing tutorial creation beyond a single domain \cite{truong2021automatic, kim2014crowdsourcing, weir2015learnersourcing}.} \DIFadd{It adopts a holistic approach by distilling common tutorial components from existing work, presenting methodologies to identify, evaluate, and assemble AI models to extract components, and introducing a guided workflow for users to inspect and modify extraction results. In this section, we reflect on the lessons learned from our exploration and discuss the broader implications.} 
\DIFdel{for future tutorial research. }


 \subsection{Selecting models and constructing pipelines}\DIFadd{~}\label{sec:futureModel}
\DIFadd{We demonstrated how to identify, evaluate, and assemble computational models into integrated pipelines to extract tutorial components. Given the rapid advancement in AI, we acknowledge that the pipeline we select may not sustain peak performance. For example, multi-modal LLMs are equipped with vision capabilities~\mbox{
\cite{gpt4v, zhang2023video, maaz2023video}}
, and dense video captioning models may improve rapidly by benefiting from large-scale pre-trained models~\mbox{
\cite{zhu2022end}}. 
Despite technological advances, our work provides enduring insights that transcend the specific models. We propose the following guidelines for future endeavors that incorporate AI into tutorial creation:
}

\begin{itemize}
    \item \textbf{\DIFadd{Adopt a multi-modal perspective}}\DIFadd{: Models across different modalities could achieve similar goals, e.g., object detectors based on video frames and LLM prompting based on transcripts can both identify object names, and each has its SoTA models. By assembling multiple pipelines with the same objective, we can explore the solution space more comprehensively without premature commitment. 
    %
    }\item \textbf{\DIFadd{Leverage strong models for cross-modal enhancement:}}
    \DIFadd{Currently, an LLM perform the best at extracting object names. Starting with the best results in one modality, we can minimize errors in other modalities, e.g., object names extracted by an LLM will guide open-vocabulary object detectors to localize objects. Future research should keep monitoring SoTA methods in different modalities.
    }\item \textbf{\DIFadd{Focus on user-centric model selection:}} \DIFadd{While each ML problem has standard metrics for evaluation, higher scores do not equate to better user experience. Though comparing models across modalities may not be straightforward due to distinct metrics, a potential universal metric could be the user's effort required to refine the output. For example, an NLVL model DORi~\cite{rodriguez2021dori} returns higher tIOU (temporal intersection over union) than ProcNets~\mbox{
\cite{zhou2018towards} }
in video segmentation, but DORi does not observe the order of steps, leading to overlapping and reverse-ordered steps, which require additional user edits. To avoid overwhelming users, we eventually dropped the model. 
}\end{itemize} 

\subsection{\DIFadd{Designing AI-Assisted user workflows}}
\label{sec:futureUI}
We believe it is important to tailor the design of mixed-media tutorial formats for different use cases. The tutorial format in our prototype shown in Figure \ref{fig:seesaw} serves only as an example interface. The following guidelines can inform future efforts to design AI-assisted tutorial creation workflows.

\begin{itemize}
    \item \textbf{\DIFadd{Simplify tutorial creation by guiding and constraining user actions:}} \DIFadd{
    The sequential editing workflow in TutoAI is structured and domain-agnostic, following the Wizard interface design pattern
    ~\mbox{
\cite{wizardWiki}}. 
One potential benefit of this approach is that the complex task of tutorial creation is transformed into a sequence of understandable stages, where the relationships between the stages are implicitly captured. Users can thus focus on individual tasks without worrying about how to structure the overall workflow. 
The UI should also ensure the results satisfy implicit constraints (e.g., the intervals of two steps should not overlap).}

    \item \textbf{Separate content from style:}
While mixed-media tutorials are available in diverse formats, TutoAI underscores the value of separating content from style. 
 \DIFadd{
 In our prototype, the user workflow focuses on extracting accurate component information; the visual representations and interactivity of the components in the tutorial are automatically applied to the extraction results. This general approach is adaptable to any mixed-media tutorial with a predefined format. Our prototype offers multiple formats for a customized consumer experience, including a list-based view of steps and a dependency diagram (Appendix Figure 16). Future tools can provide more 
 flexibility in formatting tutorials, yet the principle of separating content from style remains valid.} 

 \item\DIFadd{\textbf{Support graceful degradation:} 
 The performance of ML models can be uncertain and unpredictable. Even though the overall performance of our pipeline is reasonable, it may be disappointing in some cases. Therefore, it is important to design a UI that supports tutorial creation when AI-powered component extraction fails. To support such graceful degradation, users must be able to interpret the extraction results and make edits easily. To facilitate this, our UI is designed for low-effort error correction, e.g., users can adjust step boundaries with a range slider. In the worst case, where the extraction result is completely wrong, users can override the results and update the component manually.} 
 
\end{itemize}

 
\subsection{Cross-domain generalization: tutorials, tools, and methodologies}\label{sec:cross-domain}
\DIFadd{TutoAI is motivated by previous work's effort to generalize mixed-media tutorial creation beyond a single domain. Reflecting on our experience, we have identified multiple interpretations of cross-domain generalization: 
\begin{itemize}
    \item \textbf{CD1: Same tutorial format, diverse domains:} a tool for creating tutorials with the same format.
    \item \textbf{CD2: Same creation experience, diverse tutorial formats and domains:} a general-purpose tool for creating tutorials with diverse formats.
    \item \textbf{CD3: Same methodologies, diverse creation experiences, tutorial formats and domains:} a set of generalized methodologies to guide the design and development of tutorial creation tools; the tools can be general-purpose or domain-specific, supporting the creation of diverse tutorial formats
\end{itemize}
It is not our intention to advocate a one-size-fits-all tutorial format (CD1), as we have discussed in Section \ref{sec:UI-design-considerations} and Section \ref{sec:futureUI}. We believe a general-purpose creation tool (CD2) can be useful, as exemplified by our prototype. Nevertheless, a general-purpose tool risks overlooking domain-specific nuances in terms of both components and ML pipelines. In TutoAI, we are not only trying to build a general-purpose tool (CD2) but also propose a set of generalized methodologies for tool builders (CD3). With advancements in AI, we demonstrate the feasibility of designing tutorial creation tools systematically. Our framework, encompassing three levels – components, models, and UIs – and the associated guidelines, is adaptable to various contexts. For example, to develop a tutorial creation tool for software instructional videos, we can standardize the components first (e.g., UI widgets, commands, data), then identify, evaluate, and assemble ML pipelines based on the guidelines outlined in Section \ref{sec:futureModel}. Though the component and model details may differ, the underlying approach remains the same. 
}


\DIFdel{Currently, we use shot boundary detectors to present diverse frames as step thumbnail candidates, independent of text descriptions. In the future, thumbnail selection could leverage the text descriptions. e.g., multi-modal video summarization methods can extract representative frames and text summaries~\mbox{
\cite{he2023align, narasimhan2021clip} }
simultaneously. In the future, they have the potential to return high-quality text-dependent representative frames.
}

\DIFdel{We match objects to steps and assume steps }  \subsection{Limitations and future work}
\bpstart{Domain limitations} \DIFadd{Though TutoAI is a cross-domain framework, it does not apply to all instructional videos. Chang et al.~\mbox{
\cite{chang2021rubyslippers}} classified instructional videos into a quadrant along an object-action coordinate system, distinguishing between ``Diverse objects and diverse actions'' (cooking, car repair, makeup, etc.), ``diverse objects and few actions'' (crafts and packing, etc.), ``few objects and few actions'' (drawing, musical instrument, etc.), and ``few objects and diverse actions'' (dance, exercise, etc.). TutoAI focuses on physical tasks that involve diverse objects. For instructional videos with few objects or without concrete objects (e.g., lecture videos), TutoAI will have difficulty in constructing dependencies, as the dependency parser assumes steps} share the same object depending on each other. \DIFadd{Another related limitation is that if} the same object was referred to differently, e.g., in the berry cake video, the creator uses ``berries'' to refer to both strawberries and blueberries in the late stage, and our method fails to detect the dependency between steps containing ``berries'' and ``strawberries'' (or ``blueberries''). Future work could investigate \DIFadd{identifying abstract items and} more intelligent dependency parsing, especially dependencies between abstract concepts.

\bpstart{Representative frames selection} Currently, we use shot boundary detectors to present diverse frames as step thumbnail candidates, independent of text descriptions. In the future, thumbnail selection could leverage the text descriptions. e.g., multi-modal video summarization methods can extract representative frames and text summaries~\mbox{
\cite{he2023align, narasimhan2021clip} }
simultaneously, having the potential to return high-quality text-dependent representative frames.

\bpstart{Framework evaluation} \DIFadd{We use user-perceived component quality as a proxy for learning effects, though the two may not be positively correlated. Further research is necessary to study 
if user rating of tutorials directly translates to better learning outcomes. Besides, the fact that users interacted with TutoAI but only looked at static YouTube Chapters' screenshots may also cause bias in users' ratings.}




%% file: 10-conclusion.tex
\section{conclusion}
Transforming linear instructional videos into more browsable mixed-media tutorials will significantly elevate the learning experience, however, existing methods do not harness the full potential of the latest AI advances and are usually limited to specific domains. In response, we introduced TutoAI, a cross-domain framework for AI-assisted mixed-media tutorial creation. TutoAI provides a taxonomy for mixed-media tutorial components, a methodology to evaluate and select models for component extraction, and guidelines for UI implementation. Our empirical evaluation underscored the capability of TutoAI in extracting high-quality mixed-media tutorial components and helping authors create mixed-media tutorials. Moving forward, we believe the TutoAI framework will provide a strong foundation for future mixed-media tutorial development.